\documentclass[conference]{IEEEtran}
\IEEEoverridecommandlockouts
\usepackage{cite}
\usepackage{amsmath,amssymb,amsfonts}
\usepackage{algorithmic}
\usepackage{graphicx}
\usepackage{textcomp}
\usepackage{xcolor}
\usepackage{hyperref}
\usepackage{url}
\usepackage{subcaption}
\usepackage{multirow}
\usepackage{circuitikz}

\usepackage{xspace}

\newcommand*\mean[1]{\bar{#1}}

\begin{document}

\title{STOP! Camera Spoofing via the in-Vehicle IP Network}

\author{\IEEEauthorblockN{Dror Peri}
\IEEEauthorblockA{\textit{School of Electrical Engineering} \\
\textit{Tel Aviv University}\\
Ramat Aviv, ISRAEL \\
drorperi@mail.tau.ac.il}
\and
\IEEEauthorblockN{Avishai Wool}
\IEEEauthorblockA{\textit{School of Electrical Engineering} \\
\textit{Tel Aviv University}\\
Ramat Aviv, ISRAEL \\
yash@end.tau.ac.il}
}

\maketitle

\begin{abstract}
Autonomous driving and advanced driver assistance systems (ADAS) rely on cameras to control the driving. In many prior approaches an attacker aiming to stop the vehicle had to send messages on the specialized and better-defended CAN bus. We suggest an easier alternative: manipulate the IP-based network communication between the camera and the ADAS logic, inject fake images of stop signs or red lights into the video stream, and let the ADAS stop the car safely. We created such an attack tool that successfully exploits the GigE Vision protocol. 

Then we analyze two classes of passive anomaly detectors to identify such attacks: protocol-based detectors and video-based detectors. We implemented multiple detectors of both classes and evaluated them on data collected from our test vehicle and also on data from the public BDD corpus. Our results show that such detectors are effective against naive adversaries, but sophisticated adversaries can evade detection.

Finally, we propose a novel  class of active defense mechanisms that randomly 
adjust camera parameters during the video transmission, and verify that the received images obey the requested adjustments. Within this class we focus on a specific implementation, the width-varying defense, which randomly modifies the width of every frame. Beyond its function as an anomaly detector, this defense is also a protective measure against certain attacks: by distorting injected image patches it prevents their recognition by the ADAS logic. We demonstrate the effectiveness of the width-varying defense through theoretical analysis and by an extensive evaluation of several types of attack in a wide range of realistic road driving conditions. The best the attack was able to achieve against this defense was injecting a stop sign for a duration of 0.2 seconds, with a success probability of 0.2\%, whereas stopping a vehicle requires about 2.5 seconds. 

\end{abstract}

\maketitle


\section{Introduction}
The security of autonomous vehicles, or those with Advanced Driver Assistance Systems (ADAS), is a key component of the automotive industry. There are many published attacks against vehicles, which can be categorized into two main approaches: direct attacks accessing the CAN bus inside the vehicle and issuing commands directly (cf.\ \cite{Bozdal2020}, \cite{Avatefipour2018}), or external attacks manipulating the vehicle sensors (camera \cite{petit2015remote}, \cite{yan2016can}, \cite{nassi2021spoofing}, LIDAR \cite{cao2019adversarial}, or radar \cite{Komissarov2023}) from outside the vehicle. 

We consider a financially-motivated attacker, whose goal is to carefully stop the vehicle (e.g., to steal it or rob its cargo), as opposed to a terrorist or military adversary whose goal is to damage the vehicle, or harm the driver, passengers, or other vehicles. Direct attacks through the CAN pose two challenges to such an attacker: first, modern vehicles segregate the CAN bus from the more exposed vehicle networks using filtering CAN gateways \cite{kim2014gateway}, making the CAN bus more difficult to access; and second, safely stopping a vehicle is a complex task that requires deep understanding of the operation of multiple ECUs (brakes, gear, engine control, etc.), together with the dynamic driving conditions. Conversely, an external attack manipulating the sensors from outside the vehicle poses logistical challenges, requiring placing specialized attack equipment at pre-planned locations along the vehicle's path. 

In this paper we consider an easier attack path. We assume that the attacker can compromise a more exposed ECU, and through it gain access to the less protected in-vehicle IP-based Ethernet network. From such a vantage point, the attacker can manipulate the network traffic between the camera and the ADAS controller, injecting spoofed image patches into the video stream---and allowing the vehicle's sophisticated ADAS logic to safely stop the vehicle.

We first analyze the common GigE Vision camera-to-controller network protocol \cite{AIA_GigE_Vision}, and build an attack tool to exploit it. Our tool successfully injects images, or parts of images, into the video stream. By injecting image patches of stop signs or red traffic lights, we can get the ADAS logic to identify the traffic sign, and stop the vehicle. 

Then we analyze methods to identify such attacks and evaluate their capabilities. We first investigate two classes of \textit{passive} anomaly detectors: protocol-based detectors and video-based detectors. Protocol-based detectors look for anomalies in the packet and frame IDs, timestamps, etc., while video-based detectors look for anomalies in the image color distribution or optical flow. We implemented multiple detectors of both classes and evaluated them on data collected from self-gathered road driving scenarios in a test vehicle, and also on data from the public BDD corpus \cite{Yu_2020_CVPR}. Our results show that such passive detectors are effective against naive adversaries, however slightly more sophisticated adversaries can evade detection quite easily.

Finally, we suggest novel class  of \textit{active} defense mechanisms that randomly adjusts camera parameters during the video transmission, and verifies that the received images obey the requested adjustments. Our main choice of parameter to adjust is the frame width: By modifying each frame width by a few pixels, the active defense not only enables detection but also serves as a protective measure in some attack scenarios, since a mismatch between the requested frame width and an injected patch distorts the embedded stop sign/red light image, causing the ADAS logic to not recognize the sign and not stop the vehicle. We implemented the width-varying defense, analyzed its theoretical performance, and demonstrated its effectiveness as both an anomaly detector and a protective measure across various attack scenarios using stop sign images from the Mapillary traffic sign dataset \cite{ertler2020mapillary} in a wide range of realistic road driving scenarios.

We plan to make our data and software freely available to the research community.

\subsection{Related Work}

\subsubsection{CAN bus attacks and defenses}
Software attacks against vehicles have been widely publicized starting from the original works of \cite{Koscher2010,Checkoway:2011:CEA:2028067.2028073,miller2015remote}. These attacks succeeded to control a wide range of the automotive functions, such as disabling the brakes, stopping the engine, and so on. Since then, many attacks against vehicles followed the same approach, accessing the CAN bus inside the vehicle and issuing commands directly (cf.~\cite{Bozdal2020,Avatefipour2018}). 
 
One suggested approach to secure the CAN bus was to add some authentication to the messages on the bus by using a cryptographic Message Authentication Code (MAC) (cf.~\cite{glas,ziermann2009can+}). A similar approach was adopted by the AUTOSAR standard as defined by the Secure Onboard Communication (SecOC) mechanism \cite{SecOC}, to add some authentication and replay prevention to the vehicle's internal networks.

A different approach to try and destroy non-legitimate spoofed messages, by transmitting an \textit{active-error} flag, was suggested by Matsumoto et al.~\cite{matsumoto2012method} and Dagan and Wool~\cite{daganparrot}.

The approach that seems to be in most common use is that of a secure gateway for protecting the CAN bus network \cite{webSemconSg,kim2014gateway}.

\subsubsection{Sensor attacks and defenses}

An alternative attack approach against vehicles involves externally manipulating its sensors, from another vehicle, from a roadside vantage point, or even from a flying drone. Several works attack the vehicle through its camera, cf.~\cite{petit2015remote,yan2016can,zolfi2021translucent,nassi2021spoofing}, by placing or projecting images in front of the camera to cause attacker-desired actions such as stopping the car. 
These also include attacks aimed at computer vision algorithms in the ADAS systems that consume the camera input, such as object detection and tracking capabilities \cite{akhtar2021advances}.
Other sensors that were attacked include the LIDAR \cite{cao2019adversarial} and the FMCW radar \cite{Komissarov2023}. 

\subsubsection{WiFi and IP-based attacks} \label{sec:WifiIPattacks}
The increasing integration of wireless technologies in modern vehicles has led to growing concerns about WiFi and IP-based attacks. These cyber threats exploit vulnerabilities in a vehicle's wireless interfaces, potentially allowing unauthorized access to or control over critical vehicle systems. Research has demonstrated various attack vectors, including the deployment of rogue access points \cite{RougeAccessPoint}, the installation of malicious applications on victims' smartphones \cite{6894181} and exploitation of web vulnerabilities in the manufacturer's portal \cite{KiaHacked}. Through these methods, attackers may access the in-vehicle IP-based networks we are discussing in this paper.

\subsection{Adversary Model} \label{sec:AdversaryModel}

We consider a financially-motivated attacker, whose goal is to carefully stop the vehicle (e.g., to steal it or rob its cargo), as opposed to a terrorist or military adversary whose goal is to damage the vehicle, or harm the driver, passengers, or other vehicles.

We assume that the attacker has no access to the CAN bus. Furthermore, the attacker does not have physical access to the vehicle and does not place any roadside equipment.

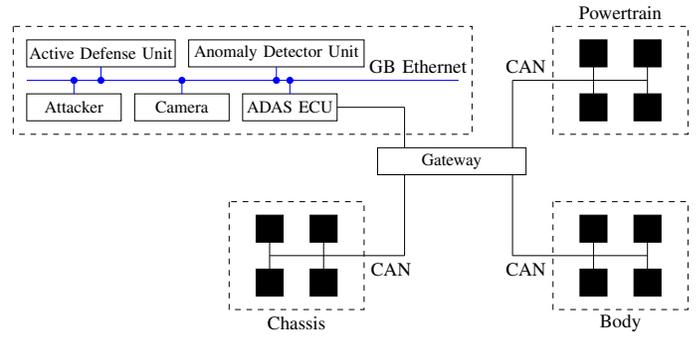
\begin{figure}[t]
\centering
\resizebox{0.5\textwidth}{!}{%
\begin{circuitikz}
\tikzstyle{every node}=[font=\small]
\draw [ line width=0.2pt ] (7.25,8.25) rectangle  node {\small Gateway} (10,7.75);
\node [font=\small] at (10.5,6.75) {};
\draw [, dashed] (10.5,7.25) rectangle  (13,5.25);
\draw [short] (9.75,6.25) -- (12.25,6.25);
\draw [ fill={rgb,255:red,0; green,0; blue,0} ] (11,7) rectangle (11.5,6.5);
\draw [short] (11.25,6.5) -- (11.25,6);
\draw [ fill={rgb,255:red,0; green,0; blue,0} ] (11,6) rectangle (11.5,5.5);
\draw [ fill={rgb,255:red,0; green,0; blue,0} ] (12,7) rectangle (12.5,6.5);
\draw [short] (12.25,6.5) -- (12.25,6);
\draw [ fill={rgb,255:red,0; green,0; blue,0} ] (12,6) rectangle (12.5,5.5);
\node [font=\small] at (10.5,10) {};
\draw [, dashed] (10.5,10.5) rectangle  (13,8.5);
\draw [short] (9.75,9.5) -- (12.25,9.5);
\draw [ fill={rgb,255:red,0; green,0; blue,0} ] (11,10.25) rectangle (11.5,9.75);
\draw [short] (11.25,9.75) -- (11.25,9.25);
\draw [ fill={rgb,255:red,0; green,0; blue,0} ] (11,9.25) rectangle (11.5,8.75);
\draw [ fill={rgb,255:red,0; green,0; blue,0} ] (12,10.25) rectangle (12.5,9.75);
\draw [short] (12.25,9.75) -- (12.25,9.25);
\draw [ fill={rgb,255:red,0; green,0; blue,0} ] (12,9.25) rectangle (12.5,8.75);
\draw (9.75,9.5) to[short] (9.75,8.25);
\draw [](9.75,6.25) to[short] (9.75,7.75);

\node [font=\small] at (4.5,6.75) {};
\draw [, dashed] (4.5,7.25) rectangle  (7,5.25);
\draw [ fill={rgb,255:red,0; green,0; blue,0} ] (5,7) rectangle (5.5,6.5);
\draw [short] (5.25,6.5) -- (5.25,6);
\draw [ fill={rgb,255:red,0; green,0; blue,0} ] (5,6) rectangle (5.5,5.5);
\draw [ fill={rgb,255:red,0; green,0; blue,0} ] (6,7) rectangle (6.5,6.5);
\draw [short] (6.25,6.5) -- (6.25,6);
\draw [ fill={rgb,255:red,0; green,0; blue,0} ] (6,6) rectangle (6.5,5.5);

\draw [](5.25,6.25) to[short] (7.75,6.25);
\draw [](7.75,6.25) to[short] (7.75,7.75);
\draw [ color={rgb,255:red,0; green,0; blue,255}, ](0.75,9.5) to[short] (8.75,9.5);
\draw [, line width=0.2pt ] (4.75,9.25) rectangle  node {\small ADAS ECU} (6.5,8.75);
\draw [, line width=0.2pt ] (2.75,9.25) rectangle  node {\small Camera} (4.5,8.75);
\draw [, line width=0.2pt ] (0.75,9.25) rectangle  node {\small Attacker} (2.5,8.75);
\draw [, line width=0.2pt ] (3.75,10.25) rectangle  node {\small Anomaly Detector Unit} (7,9.75);
\draw [, line width=0.2pt ] (0.75,10.25) rectangle  node {\small Active Defense Unit} (3.5,9.75);
\draw [, dashed] (0.5,10.5) rectangle  (9,8.5);
\draw [ color={rgb,255:red,0; green,0; blue,255}, short] (1.625,9.25) -- (1.625,9.5);
\draw [ color={rgb,255:red,0; green,0; blue,255}, short] (3.625,9.25) -- (3.625,9.5);
\draw [ color={rgb,255:red,0; green,0; blue,255}, short] (5.625,9.25) -- (5.625,9.5);
\draw [ color={rgb,255:red,0; green,0; blue,255}, short] (2.125,9.75) -- (2.125,9.5);
\draw [ color={rgb,255:red,0; green,0; blue,255}, short] (5.375,9.75) -- (5.375,9.5);
\node at (1.625,9.5) [circ, color={rgb,255:red,0; green,0; blue,255}] {};
\node at (2.125,9.5) [circ, color={rgb,255:red,0; green,0; blue,255}] {};
\node at (3.625,9.5) [circ, color={rgb,255:red,0; green,0; blue,255}] {};
\node at (5.625,9.5) [circ, color={rgb,255:red,0; green,0; blue,255}] {};
\node at (5.375,9.5) [circ, color={rgb,255:red,0; green,0; blue,255}] {};
\node [font=\normalsize] at (8,9.75) {GB Ethernet};
\node [font=\normalsize] at (7.5,6) {CAN};
\node [font=\normalsize] at (10,6) {CAN};
\node [font=\normalsize] at (10,9.75) {CAN};
\node [font=\normalsize] at (11.75,10.75) {Powertrain};
\node [font=\normalsize] at (5.75,5) {Chassis};
\node [font=\normalsize] at (11.75,5) {Body};
\draw [](6.5,9) to[short] (7.75,9);
\draw [](7.75,9) to[short] (7.75,8.25);

\end{circuitikz}
}%

\caption{Networking Architecture Structure. Note the attacker's position in an ECU connected to the GB Ethernet.}
\label{fig:ethernet_scheme}
\end{figure}

We assume the attacker has network access by exploiting a vulnerability in an ECU connected to the vehicular Ethernet (e.g., \cite{RougeAccessPoint,6894181,mazloom2016security,lin2019common}), see Figure~\ref{fig:ethernet_scheme}. From this vantage point the attacker is able to mount typical IP- and TCP/UDP-based network attacks such as ARP-poisoning, IP spoofing and TCP session hijacking. We assume that the traffic between the camera and the ADAS logic is transmitted using IP-based protocols over this Ethernet. Therefore, within this setup, the attacker is able to achieve an in-path Man-in-the-Middle (MitM) network position between the camera and ADAS logic.

Even though modern vehicles often employ multiple sensors and cameras, sensor fusion does not mitigate our attack targeting a single camera. This stems from the ``safety first'' principle followed by many autonomous and semi-autonomous vehicles. Under this principle, the ADAS logic is programmed to respond to high-confidence detections from even a single sensor, prioritizing potential safety risks over the need for multi-sensor coherence, as was demonstrated successfully by~\cite{nassi2020phantom}.

When we discuss defense mechanisms we assume that the defender's anomaly detection module is also connected to the Ethernet, either in parallel or directly in front of the ADAS logic controller, as illustrated in Figure~\ref{fig:ethernet_scheme}.
We do not assume the existence of any cryptographic controls: e.g., no shared secret between the ADAS and camera, and no authenticated (e.g., VPN-type) connection between them. Our goal is to mitigate the attacks using existing camera and ADAS software and hardware.

When discussing attacks against both active and passive defense mechanisms, we assume the attacker must obey real-time constraints, and is non-adaptive. The attacker is able to capture the \textit{previous} frame, modify parts of it with pre-prepared data, and inject the modified packets into the current frame's stream. However, the real-time constraint implies that the attacker cannot capture the \textit{current} frame from the camera (and block it from reaching the ADAS), dynamically manipulate it in memory, fragment it into packets and transmit them in time. The non-adaptive constraint implies that the attacker has a MitM position on the camera-to-ADAS video channel, but not on the camera's control channel.

\subsection{Contributions}

We summarize our contributions as follows:
\begin{itemize}
    \item Analysis of possible spoofing attacks targeting the GigE Vision protocol.
    \item Development of an attack tool which successfully implements various manipulations against GigE Vision, including full frame injection, stripe injection and patch injection.
    \item Design of multiple passive anomaly detection methods utilizing both constant metadata and varying metadata of the transmission, and utilizing image processing techniques applied to the raw video stream.
    \item Implementation of the above anomaly detectors, and evaluation of their performance on data collected from road driving scenarios in a test vehicle, and also on data from the public BDD100k corpus \cite{Yu_2020_CVPR}.
    \item Design of a novel class of active defenses, and in particular the implementation and evaluation of the width-varying defense. 
    \item Rigorous analysis and extensive experimentation of the width-varying defense in realistic road driving scenarios, demonstrating its ability to detect and protect the ADAS logic from full-frame, stripe and patch injection attacks.
\end{itemize}

All our defense mechanisms, both passive and active, can be utilized
without changing the ADAS logic or the camera. 
In fact, they can even be added to the vehicle as an after-market component, 

\section{Preliminaries}
\subsection{The GigE Vision Protocol}
The GigE Vision protocol \cite{AIA_GigE_Vision} is a standard interface providing high-speed video and related control data over Ethernet networks. GigE has two main sub-protocols running over UDP: the control protocol (GVCP) defining how to control and configure the camera; and the stream protocol (GVSP) defining how images are transferred.

\begin{figure}[t]
\centerline{\includegraphics[width=0.5\textwidth]{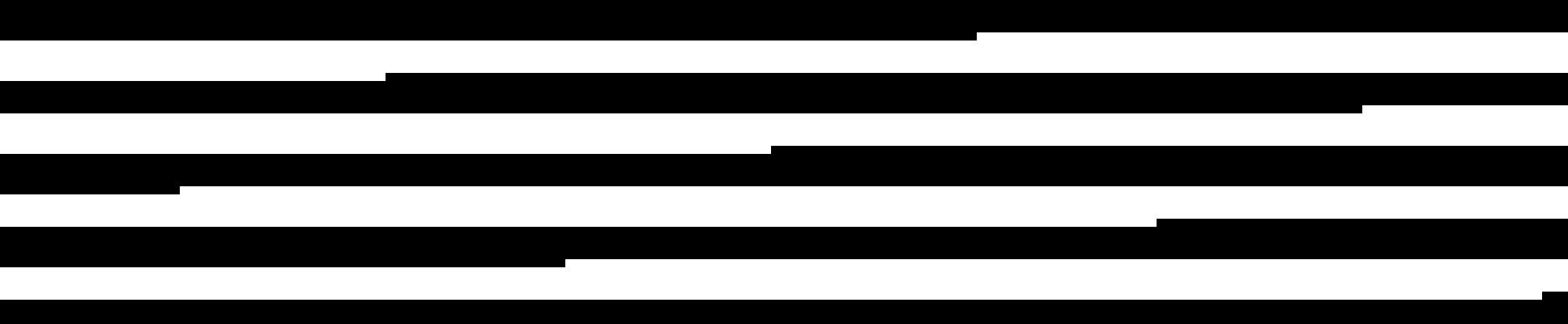}}
\caption{
A schematic representation of the first 40 rows in a frame, with 1936 pixels per row, each pixel encoded by one byte, and 9000-byte GVSP packets. The pixels of successive packets alternate between black and white.}
\label{fig:gvsp_framing}
\end{figure}

The standard indicates that the camera listens on UDP port 3956 to receive GVCP commands. During the initial phase of GVCP control communication, a discovery process occurs, which involves determining the port numbers to be used by GVSP. 
As part of the negotiation the ADAS logic informs the camera to which UDP port to transmit the GVSP packets, and queries what would be the source port that the camera will use. In our environment the GVSP packets were always sent from source port UDP/10010, while the receiving port on the ADAS logic side was a dynamically selected high port number.

Following the discovery process, the raw frames are streamed in the GVSP protocol, without compression. Due to the large size of the image frames, they are fragmented into multiple packets to be sent over the Gigabit Ethernet network: in our environment the camera used 9000-byte packets. A frame is sent as a sequence of: 
\begin{enumerate}
    \item A leader packet containing metadata such as height, width, and timestamp for the incoming image;
    \item followed by multiple payload packets containing the raw pixel values;
    \item and finally a trailer packet signaling the end of the frame.  
\end{enumerate}
Each leader packet also includes a unique sequential identification number, referred to as the block ID, and each packet within the block includes an identification number as well.
The transmitted pixels in the payload packets are sent row by row. 
In our environment the camera captured images of 1936$\times$1216 pixels, each pixel encoded by one byte. So each 9000-byte GVSP packet (including a 50 byte header) on average includes pixel values from 4.6 rows. A full frame with 1216 rows requires 263 packets, plus a header and a trailer packet. Figure~\ref{fig:gvsp_framing} illustrates the packet boundaries over the first 40 rows of a frame. 

The GigE Vision standard does not mandate checking that the block and packet IDs increase sequentially: this type of validation is left to the receiver logic. For example, in the default Allied Vision software ``vimba''
\cite{VimbaX}
the validity of IDs is not checked. On the other hand, in the MATLAB image acquisition toolbox~\cite{MatlabImageAcquisitionToolbox}, replaying frames ``from the past'' raises an error.

Controlling the camera’s parameters, such as the exposure time or encoding format, is done via GVCP requests for reading from or writing to specific registers storing the parameter values.

\subsection{The Experimental Car Setup} \label{sec:ExperimentalCar}
We used an experimental test car to gather data in a controlled environment where we can create our own dataset, analyze both the video stream and the communication metadata, and deploy our own attack and defense mechanisms. The platform is a standard passenger vehicle, a KIA Niro, equipped with a variety of sensors, including cameras, radars, and LIDAR. Through an Ethernet switch, the sensors are connected to a central PC running experimental ADAS logic, which allows inspection of the raw data transmitted between the sensors and the computer. This setup allows us to simulate an attacker who infiltrated the systems and has network access to the sensor’s data. 
The installed front-facing camera is an Allied Vision Prosilica GT1930C \cite{AlliedVision}, a 2.35 megapixel camera with a GigE Vision compliant Gigabit Ethernet port. The communication between the camera and the PC follows the GigE Vision protocol.

The Prosilica camera in our experimental car encodes images using the Bayer encoding \cite{bayer1976color}. This encoding is a technique for generating color images utilizing a single imaging sensor, rather than three individual sensors for the red, green, and blue components. The sensor is covered by a color filter array (CFA) with a mosaic color pattern such that each $2\times 2$ box of pixels consists of 2 green values (on the diagonal), 1 red value, and 1 blue value. To obtain a full-color image, various demosaicing algorithms can be used to interpolate the values for a specific pixel. For our purposes, and especially when we discuss the width-varying defense in Section~\ref{sec:ActiveDefense} this implies that both the frame width and height must be even.

To collect data, we drove the car in urban environments and on highways, both during the day and at night, while sniffing the communication between the camera and the computer.

\begin{figure}[t]
    \centering
    \begin{subfigure}[t]{0.5\textwidth}
        \centering
        \resizebox{\textwidth}{!}{%
        \begin{circuitikz}
        \tikzstyle{every node}=[font=\LARGE]
        \draw [ line width=1pt ] (4,10.25) rectangle (5,5);
        \node [font=\LARGE] at (4.5,10.75) {Camera};
        \node [font=\LARGE] at (0,10.75) {Attacker};
        \draw [, line width=1pt ] (-0.5,10.25) rectangle (0.5,5);
        \draw [, line width=1pt ] (8.5,10.25) rectangle (9.5,5);
        \node [font=\LARGE] at (9,10.75) {ADAS};
        \node [font=\normalsize] at (6.75,9.5) {GVSP Frame};
        \draw [line width=1pt, ->, >=Stealth] (5,9.25) -- (8.5,9.25);
        \draw [line width=1pt, ->, >=Stealth] (0.5,8.25) -- (4,8.25);
        \node [font=\normalsize] at (2,8.5) {GVCP Stop};
        \draw [line width=1pt, ->, >=Stealth] (0.5,7.75) -- (8.5,7.75);
        \draw [line width=1pt, ->, >=Stealth] (0.5,6.5) -- (4,6.5);
        \node [font=\normalsize] at (2,6.75) {GVCP Start};
        \node [font=\normalsize] at (6.75,8.75) {GVSP Frame};
        \draw [line width=1pt, ->, >=Stealth] (5,8.5) -- (8.5,8.5);
        \node [font=\normalsize] at (2,8) {GVSP Frame};
        \draw [line width=1pt, ->, >=Stealth] (0.5,7) -- (8.5,7);
        \node [font=\normalsize] at (2,7.25) {GVSP Frame};
        \node [font=\normalsize] at (6.75,6.5) {GVSP Frame};
        \draw [line width=1pt, ->, >=Stealth] (5,6.25) -- (8.5,6.25);
        \node [font=\normalsize] at (6.75,5.75) {GVSP Frame};
        \draw [line width=1pt, ->, >=Stealth] (5,5.5) -- (8.5,5.5);
        \end{circuitikz}
        }%
    \end{subfigure}\hfill
    
\caption{The full-frame injection scheme}\label{fig:attacking_prototype_scheme}
\end{figure}
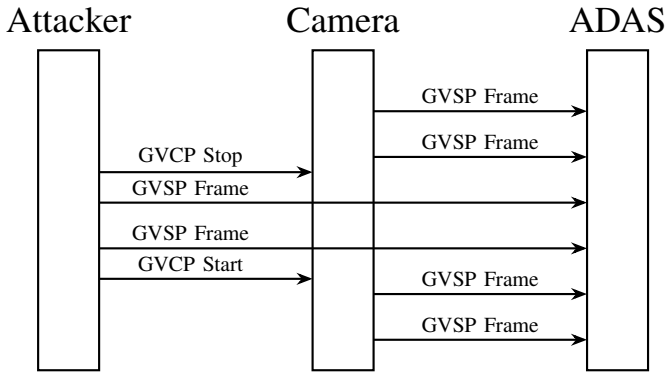

\subsection{External Datasets}
To evaluate the passive video-processing detectors, besides the data collected from our experimental car we also used an external dataset: for such detectors only the video stream itself, rather than the raw metadata, was required. In \cite{Bogdoll2023} there is a survey of perception datasets of anomaly detection in autonomous driving, and we used the BDD100k-Anomaly dataset \cite{Yu_2020_CVPR}. It is a large and diverse driving video dataset collected by a real driving platform and covers different weather conditions, as well as different times of day. It contains 100000 videos, out of which we used the 1000 videos of the test set.

To assess the effectiveness of the width-varying defense in guarding against injections, we used the Mapillary traffic sign dataset for detection \cite{ertler2020mapillary}. Specifically, we selected a subset of images which were annotated as stop sign images in the Mapillary training set and detected by MobileNet~\cite{Howard2017} as well.
We then injected stripes or patches from these images into our collected dataset. The Mapillary dataset contains sign images from all around the world, captured at various weather, season and time-of-day conditions.

\section{The Attack Tool} \label{sec:attackTool}

To demonstrate the practicality of our attack model, we implemented a prototype attack tool. Our tool is able to attain a man-in-the-middle position in the GVSP session, and implements both full-frame injection and partial frame injection. To launch a full-frame injection attack, the tool sniffs the legitimate bidirectional GigE packets sent between the camera and ADAS logic, briefly stops the legitimate transmission from the camera using a GVCP command, sends pre-prepared packets representing the fake frame to the ADAS, and reactivates the camera. A sketch of the attack tool actions is depicted in Figure~\ref{fig:attacking_prototype_scheme}.

In order to initiate this type of attack, the attacker needs to intercept the link between the camera and the ADAS logic at least once for each new video stream. This interception is necessary to configure the dynamically changing port numbers and inject packets into the receiving socket.

Within the GigE Vision protocol, the acquisition state is modifiable by issuing a GVCP write register command to a designated register. An attacker can manipulate the stream's continuity by using this command, toggling between halting and resuming the original continuous stream with values of 0 or 1, respectively. Our prototype demonstrates this attack by interrupting the transmission, sending multiple frames featuring either a stop sign or a red light, and subsequently restoring the initial link. We simulate the anticipated vehicle response to encountering a stop sign or red light by executing the MobileNetV3 \cite{Howard2017} object detection algorithm on every frame, and indicating the detection with a bounding box. We used the Scapy library to execute the attack from the computer connected to the camera. 

\begin{figure*}[t]
\centering
    \begin{subfigure}[t]{0.3\textwidth}
    \centering
    \includegraphics[width=\textwidth]{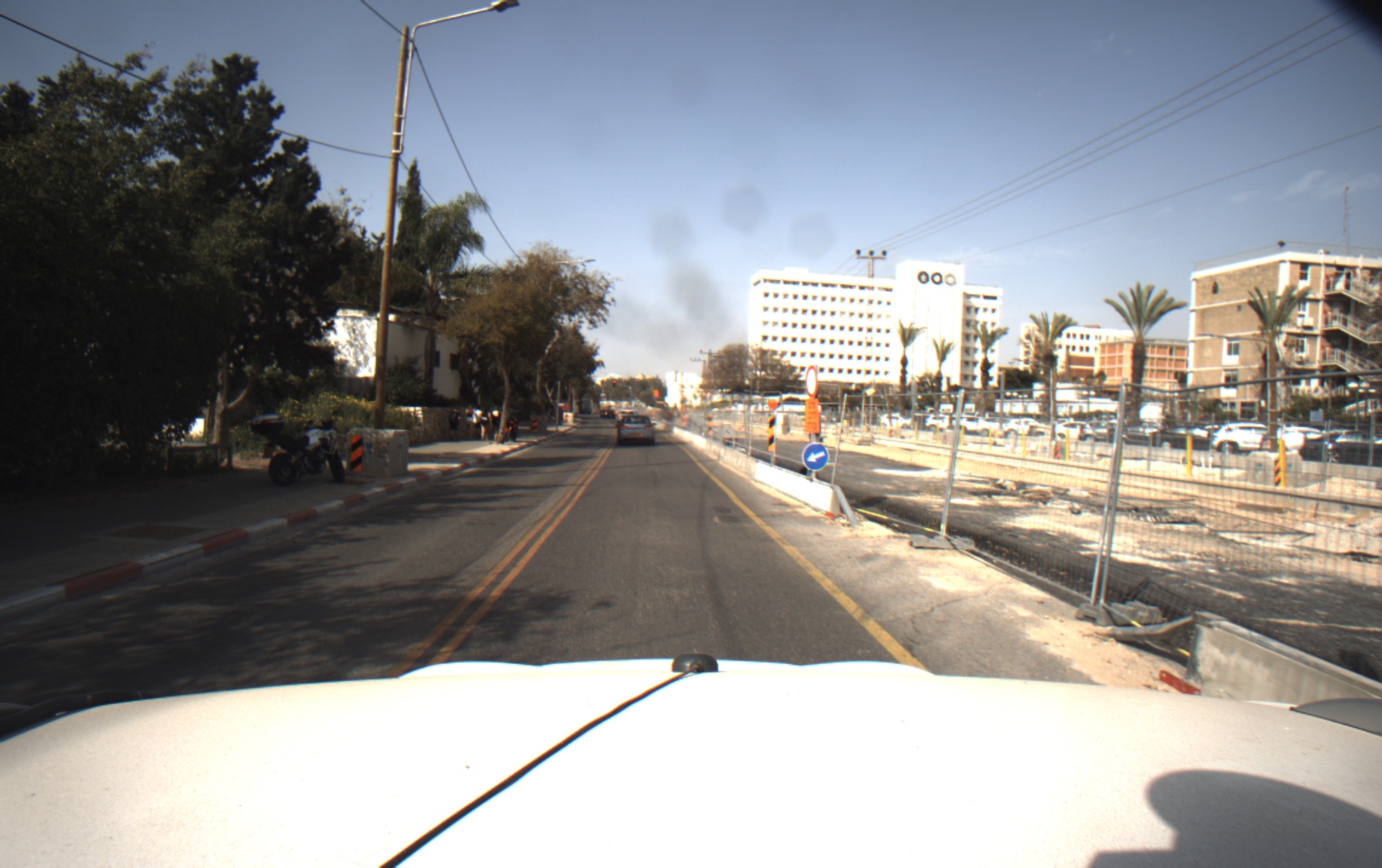}
    \caption{Before}
\end{subfigure}\hfill
\begin{subfigure}[t]{0.3\textwidth}
    \centering
    \includegraphics[width=\textwidth]{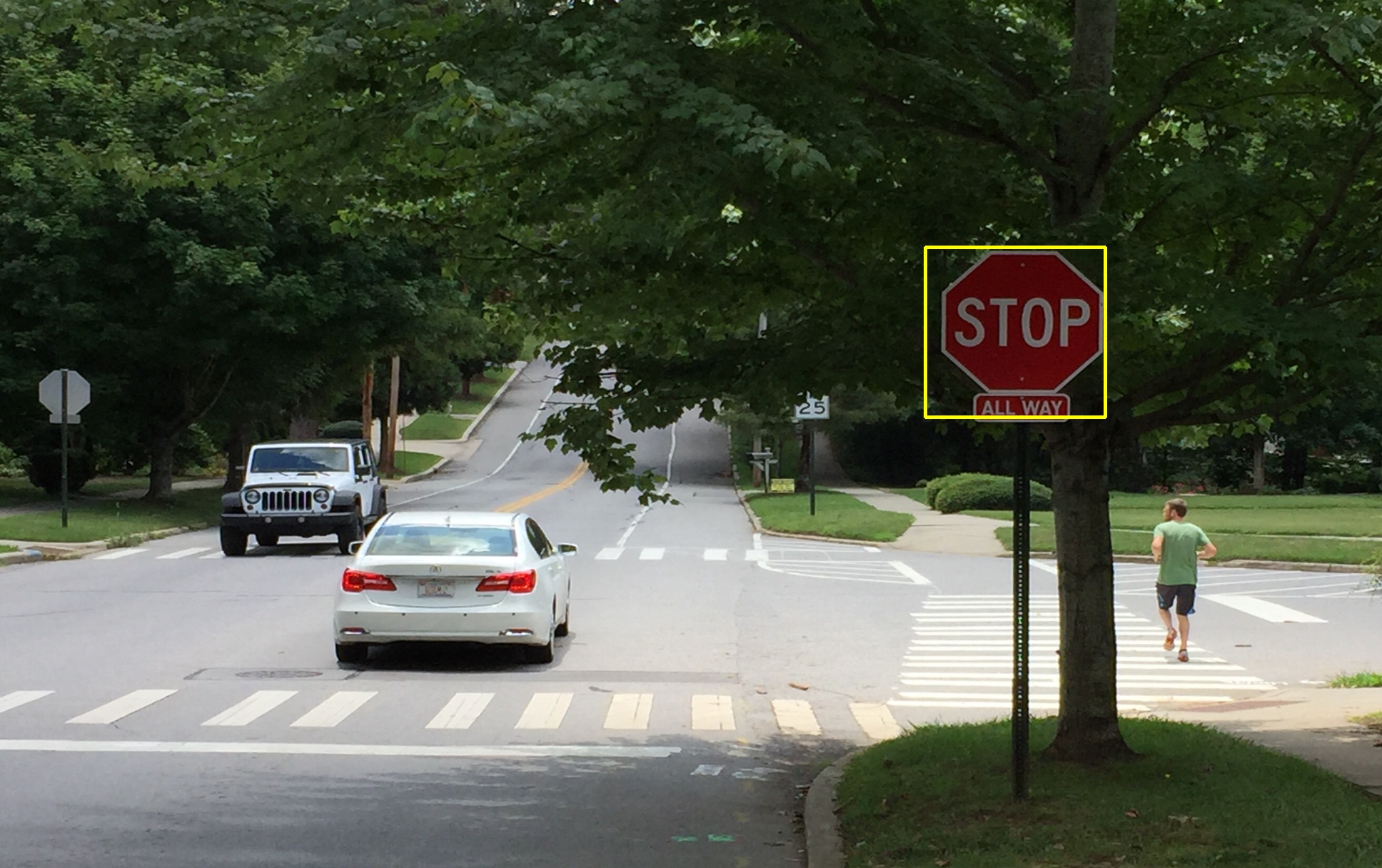}
    \caption{During}
    \label{fig:full_frame_injection}
\end{subfigure}\hfill
 \begin{subfigure}[t]{0.3\textwidth}
    \centering
    \includegraphics[width=\textwidth]{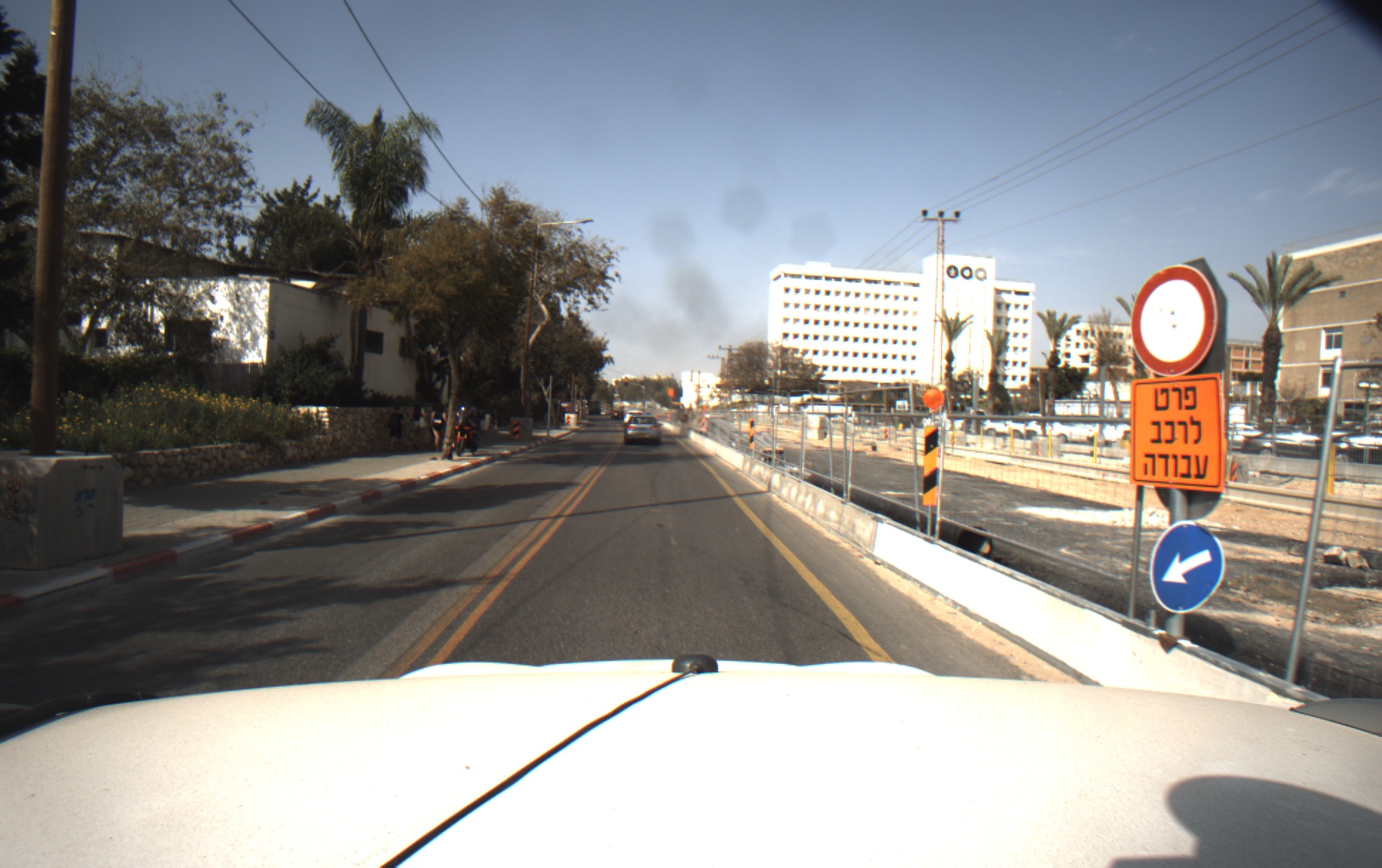}
    \caption{After}
\end{subfigure}

\caption{Full frame injection attack sequence. Note the box around the stop sign in (b) indicating that the sign was recognized.}
\label{fig:attacking_sequence}
\end{figure*}

In Figure~\ref{fig:attacking_sequence}, the stages of the full frame injection attack are depicted, showcasing frames captured before, during, and after the attack. Prior to and following the attack, the scene appears normal, but within the manipulated frame, an image featuring a stop sign is injected \cite{stopsign}. The object detector successfully identifies the stop sign, triggering the ADAS logic of the vehicle to stop the vehicle.

\begin{figure*}[t]
\centering
\begin{subfigure}[t]{0.3\textwidth}
    \centering
    \includegraphics[width=\textwidth]{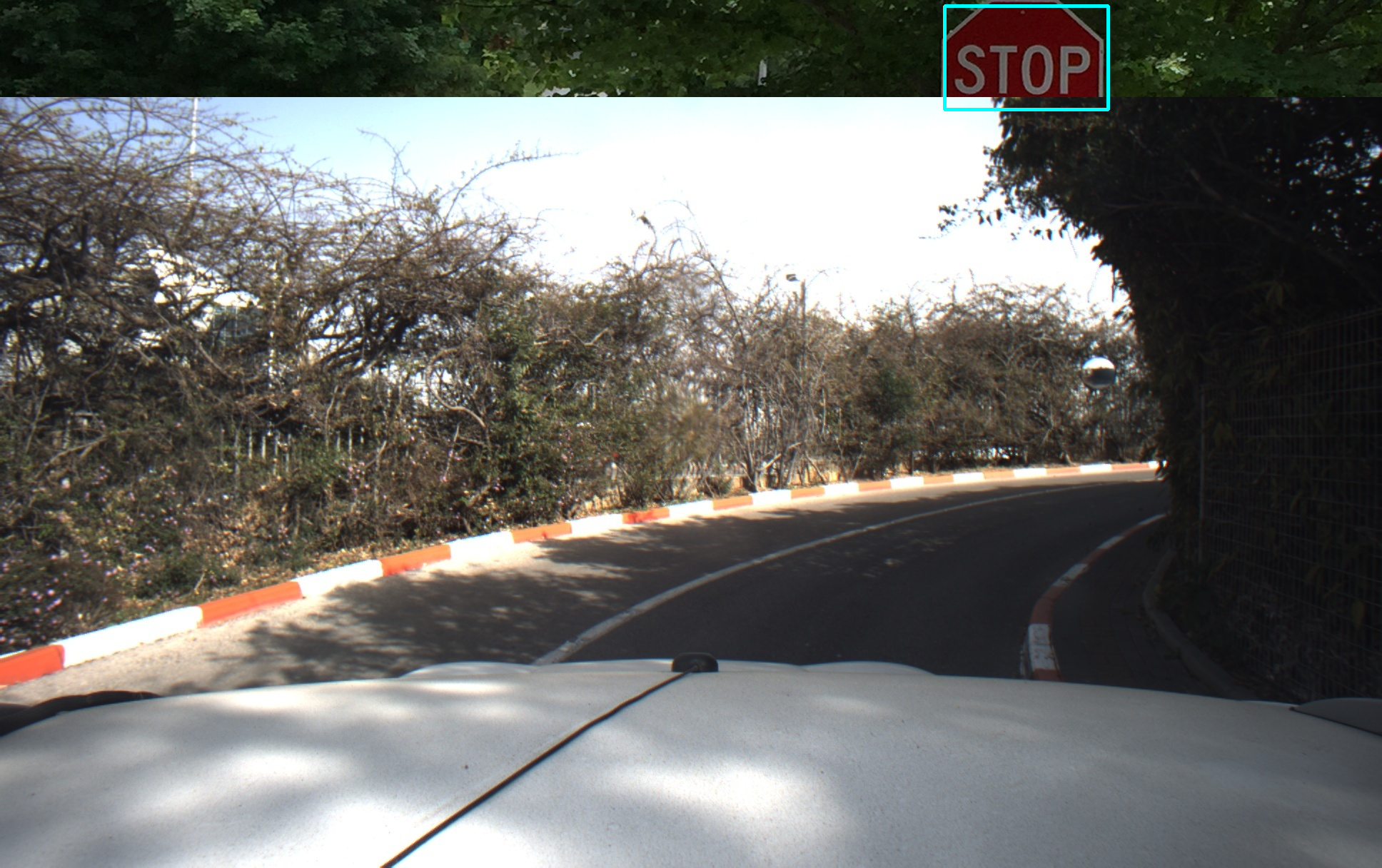}
    \caption{Stripe injection of a stop sign}
    \label{fig:stripe_injection_attack_stop_sign}
\end{subfigure}\hfill
\begin{subfigure}[t]{0.3\textwidth}
    \centering
    \includegraphics[width=\textwidth]{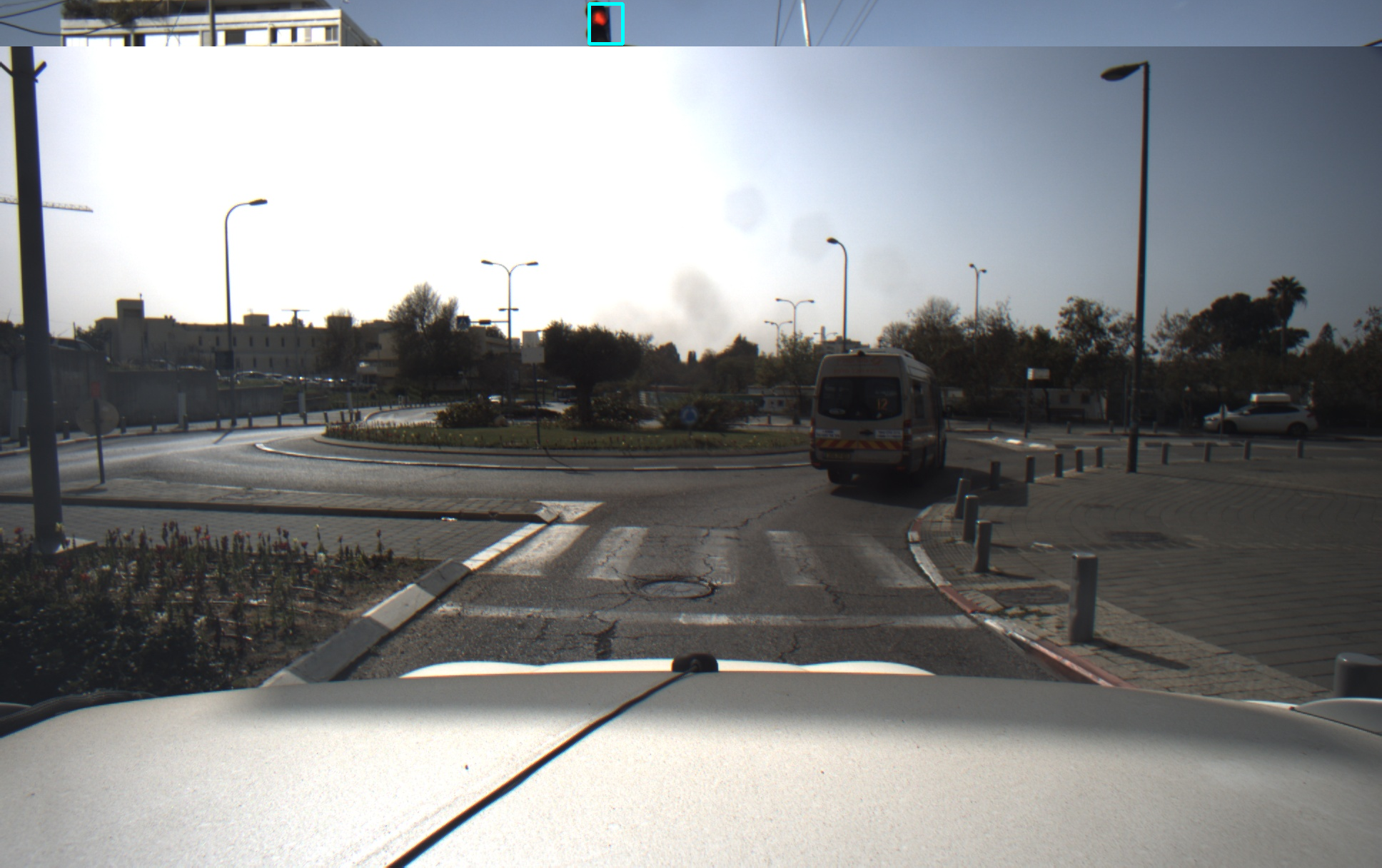}
    \caption{Stripe injection of a red light}
    \label{fig:stripe_injection_attack_red_light}
\end{subfigure}\hfill
\begin{subfigure}[t]{0.3\textwidth}
    \centering
    \includegraphics[width=\textwidth]{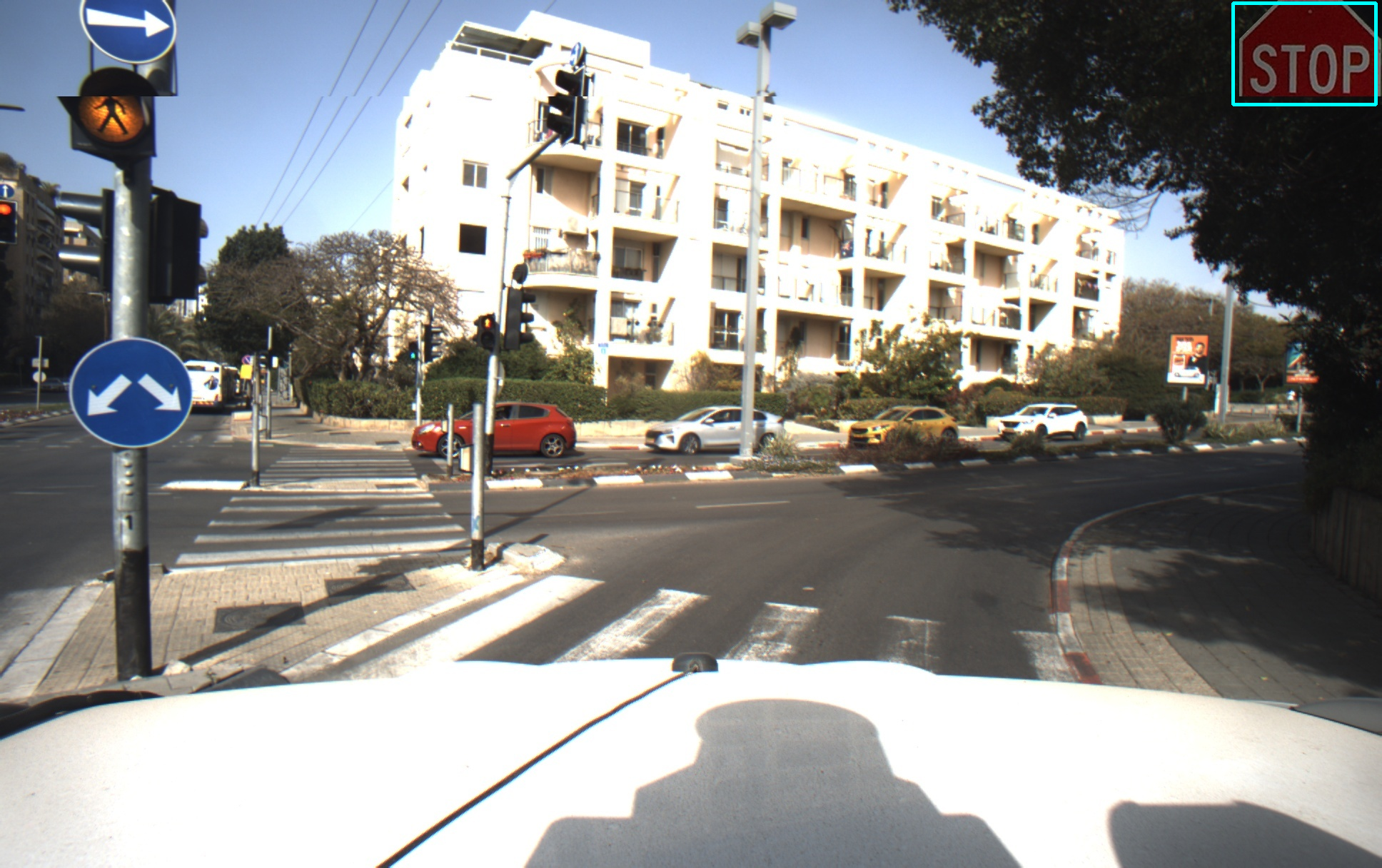}
    \caption{Patch injection of a stop sign}
    \label{fig:patch_injection_attack_stop_sign}
\end{subfigure}
\caption{Attack tool examples: the boxes around the signs or red lights indication a successful detection.}
\label{fig:attack_tool_examples}
\end{figure*}

Instead of injecting a full frame, a more stealthy attack injects a horizontal stripe, without stopping the GVSP session. Recall that in the GVSP protocol, frames are divided into multiple packets, where each packet consists of several consecutive rows. Each frame is 
assigned with an ID, and each packet is associated with both its unique packet ID and its block ID. Packets with the same block ID are part of the same frame. The attack tool successfully exploits a race condition by transmitting fraudulent packets with the correct block ID ahead of the genuine camera, which results in manipulation of a portion of the frame sent to the ADAS. In Figure~\ref{fig:stripe_injection_attack_stop_sign} and ~\ref{fig:stripe_injection_attack_red_light}, illustrations of such an attack are presented, where the traffic sign detector successfully identifies the stop sign and the red light, despite the fact that under 10\% or the frame is being manipulated. 
We call this type of injection a stripe injection.

The last injection attack, called patch injection, involves inserting only the desired traffic sign into the frame. To do so, we extract a stripe from the previous legitimate frame, insert the patch into the stripe, fragment the stripe into packets for transmission, and utilize the same race condition as in the stripe injection. Patch injection implements minimal alterations to the frame apart from the injected patch, making it very stealthy. In Figure~\ref{fig:patch_injection_attack_stop_sign}
shows an example of the attack, where a patch containing the stop sign is injected in the top right corner of the frame.

\section{Defenses}
\subsection{Passive Defense}
\subsubsection{Detector Categories}
To develop anomaly detection mechanisms, we initially characterize the possible signals within the video stream into two categories: protocol-based signals within the metadata of the stream,  and video-based perceptual signals within the stream itself.

The protocol-based metadata signals may be sufficient to detect a naive attack. In the GVSP protocol these parameters are transmitted in the leader packet of every frame and include the width and height of the frame and the color encoding of each pixel, for example BayerRG8. Each frame also includes temporally varying parameters. Examples of such parameters are timestamp and sequential ID which exist in the leader packet as well. Thus one can design anomaly detectors that flag obvious deviations such as unexpected changes in the image dimensions, non-monotone changes in the timestamps and sequence numbers.

Video-based perceptual signals involve analyzing the video stream, for which we used the OpenCV library \cite{itseez2015opencv}. We make two assumptions about a normative video stream: First, the frames exhibit variability, and second, the change between consecutive frames is expected to be low, as a high sample rate indicates limited changes to the scene alterations. Based on these assumptions many signals can be used, among which we chose to evaluate three. 

The first signal is the mean squared distance (MSE) between two consecutive frames, which aims to detect lack of variability---i.e., to detect nearly identical consecutive frames. 

The second signal is the hue-saturation histogram of the frame, where the frame is first converted to the HSV channel space, then the hue channel is binned using 50 bins and the saturation channel is binned using 60 bins and a combined histogram of the two is calculated. The histogram is a global signal of the colors of the frame where the value channel is ignored to be more robust to illumination conditions. 

The third signal aims to extract local features, and is inspired by the Lucas-Kanade optical flow estimation method \cite{lucas1981iterative,  Bouguet1999PyramidalIO}. In the implementation we used, interest points are extracted from every frame using the Shi-Tomasi corner detector\cite{Shi1994GoodFT}. Then, the optical flow is estimated using these points. The signal is the error of the corresponding features between two consecutive frames.

\subsubsection{Properties of Protocol-Based Detectors} \label{sec:networkBased}
Several protocol-based signals are expected to remain constant: they represent parameters of the camera that are chosen by the manufacturer and stored in a registry. Every deviation from the expected value should trigger an alert. Examples of constant parameters include the dimensions of the image and the color encoding.

To overcome this basic detection mechanism, the attacker simply needs to deduce the camera parameters once and adjust the attack accordingly. Additionally, there is a likelihood that vehicles of the same model and manufacturer may share identical parameters.

More dynamic signals can be derived from the varying metadata parameters. First, it is easy to predict the next frame ID in a normative session: a frame whose ID number is ``from the past'' or ``from the future'' should trigger an anomaly. To mitigate the occurrence of false alarms in situations involving packet loss, we implement a window that considers a limited range of accepted identification numbers.
Furthermore, we suggest two uses for the timestamp of the frames. The first involves utilizing knowledge of the frame rate in the stream to confirm that the difference between the timestamps of consecutive frames aligns with the expected rate, up to a certain threshold. The second approach is to compare the timestamp rates against an external clock. The presence of packet loss also poses a challenge for these approaches: if every deviation from normal behavior is treated as an anomaly, each instance of packet loss would trigger a false alarm. Both parameters can be combined by setting the expected rate to a multiple of the basic sample rate times the difference of the accepted identification numbers. 

From the perspective of an attacker, bypassing the ID-based and timestamp-based detectors require slightly more effort. Now, the attacker must eavesdrop on the communication stream and adjust the metadata parameters of the fake frames in real time, and is not able to simply inject static pre-selected images.

\subsubsection{Properties of Video-based Detectors} \label{sec:videoBased}

A simple attack might be a fixed-image-loop attack, in which the attacker repeatedly transmits the same image. To guard against this, we suggest a minimal threshold on the mean squared distance (MSE) between two consecutive frames, 
\begin{equation}
\frac{1}{N} \sum_{p} (I_t[p] - I_{t-1}[p])^2 < \text{th}\label{mse_detector}
\end{equation}
where $N$ is the total number of pixels in a frame, $I_t$ is the current frame and $I_{t-1}$ is the previous frame. If the MSE distance is below the threshold, the two frames are considered identical, signaling an anomaly.

To alert on rapid variation of the scene we suggest two different algorithms. One is based on Bhattacharyya distance \cite{1089532}
between the hue-saturation histograms of two consecutive frames: 
\begin{equation}
\sqrt{1 - \frac{1}{\sqrt{\mean{H}_{t-1} \mean{H}_t N^2}} \sum_{k} \sqrt{H_{t-1}(k) H_t(k)}} > \text{th}
\label{histogram_detector}
\end{equation}
where $H$ is the hue-saturation histogram and $N$ is the total number of bins.
If the distance is larger than some threshold it means the two frames are too different and it is considered an anomaly.

The second is based on the error of the corresponding Lucas-Kanade points-of-interest between two consecutive frames. In a normative stream with a sufficient sample rate, most of the points-of-interest should be matched with small errors caused by the movement of the vehicle's camera. If the points-of-interest are not matched at all or if the error is too large, it suggests an abnormal video stream.  

\begin{figure}[t]
\centerline{\includegraphics[width=0.5\textwidth]{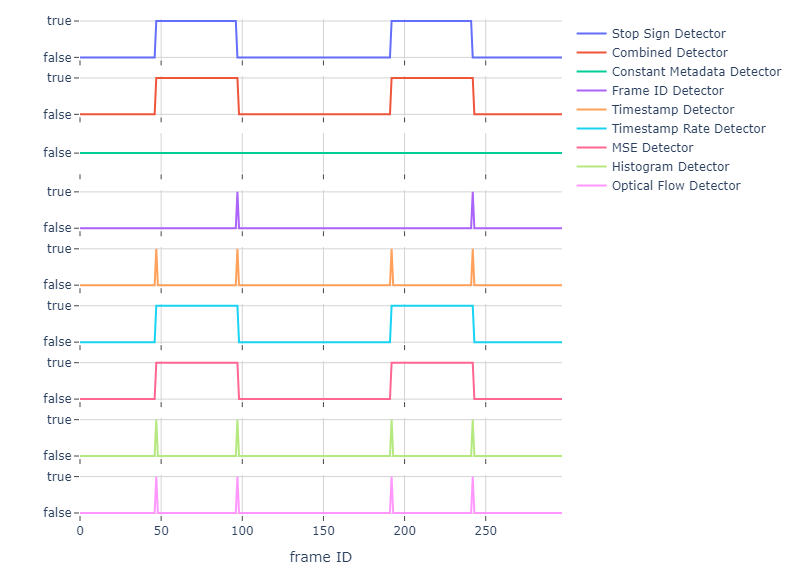}}
\caption{Anomaly detection results for the full frame injection attack.}
\label{fig:detections}
\end{figure}

\begin{figure*}[t]
\centering
    \begin{subfigure}[t]{0.3\textwidth}
    \centering
    \includegraphics[width=\textwidth]{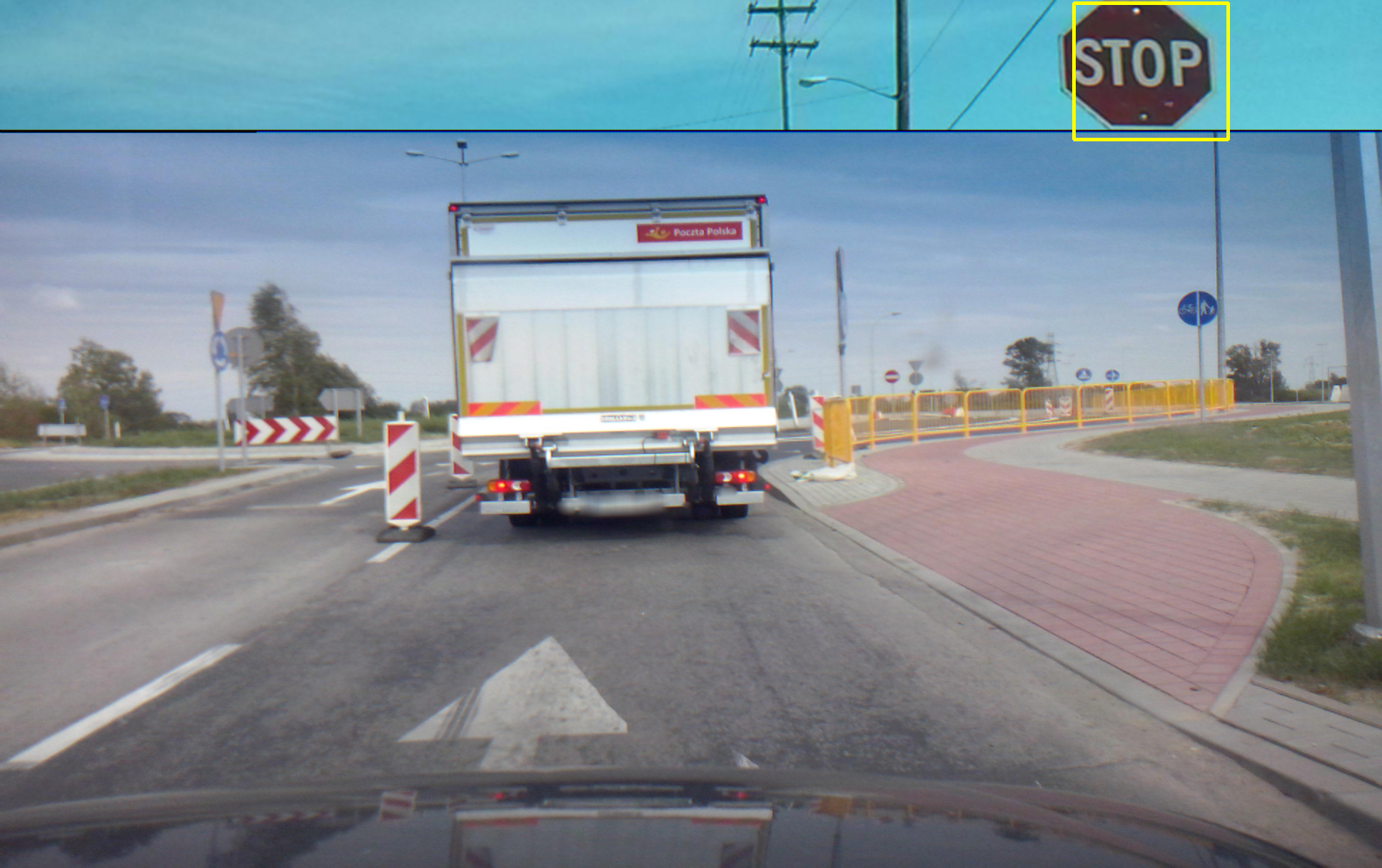}
    \caption{Width difference 0}
\end{subfigure}\hfill
\begin{subfigure}[t]{0.3\textwidth}
    \centering
    \includegraphics[width=\textwidth]{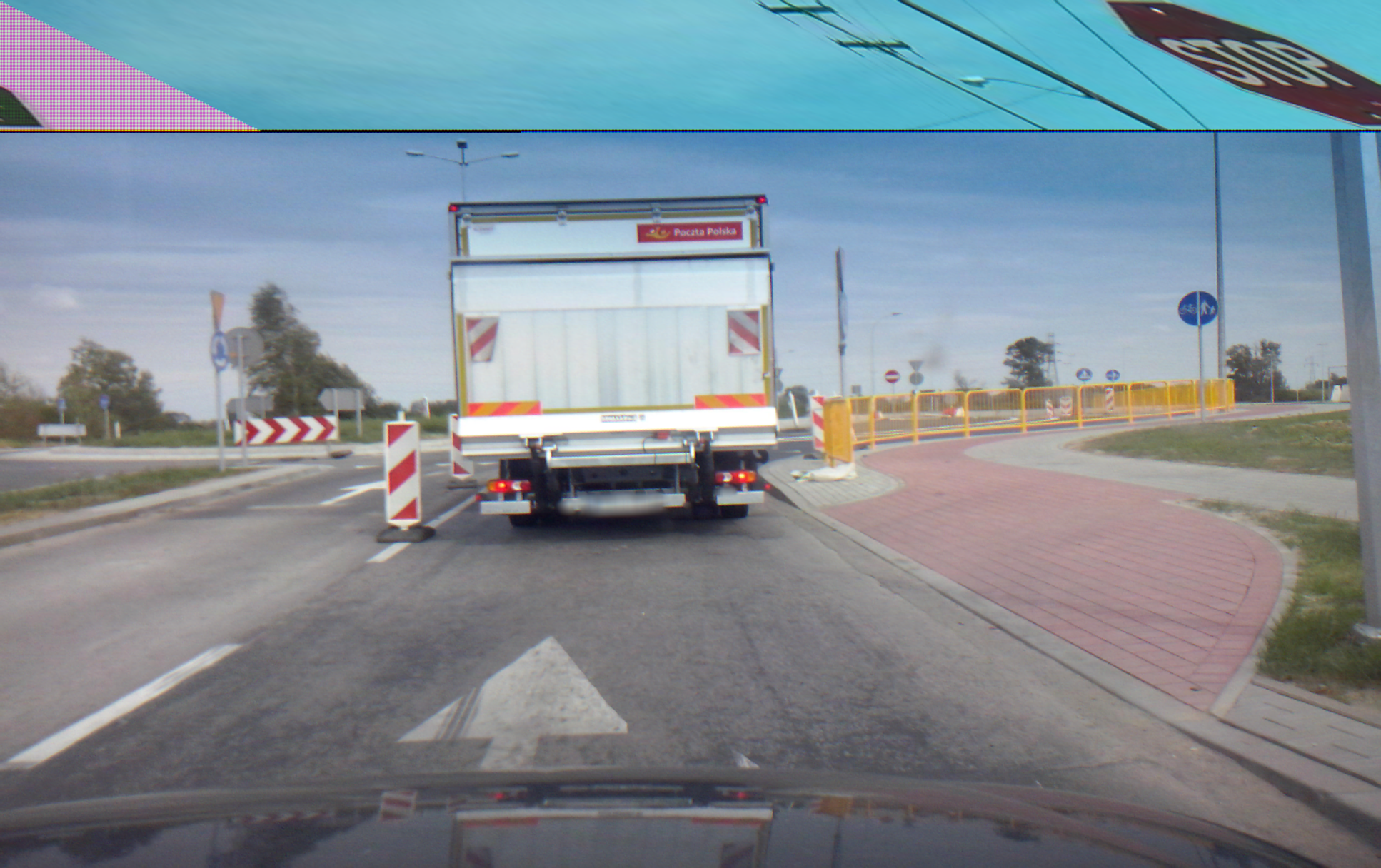}
    \caption{Width difference 2}
    \end{subfigure}\hfill
 \begin{subfigure}[t]{0.3\textwidth}
    \centering
    \includegraphics[width=\textwidth]{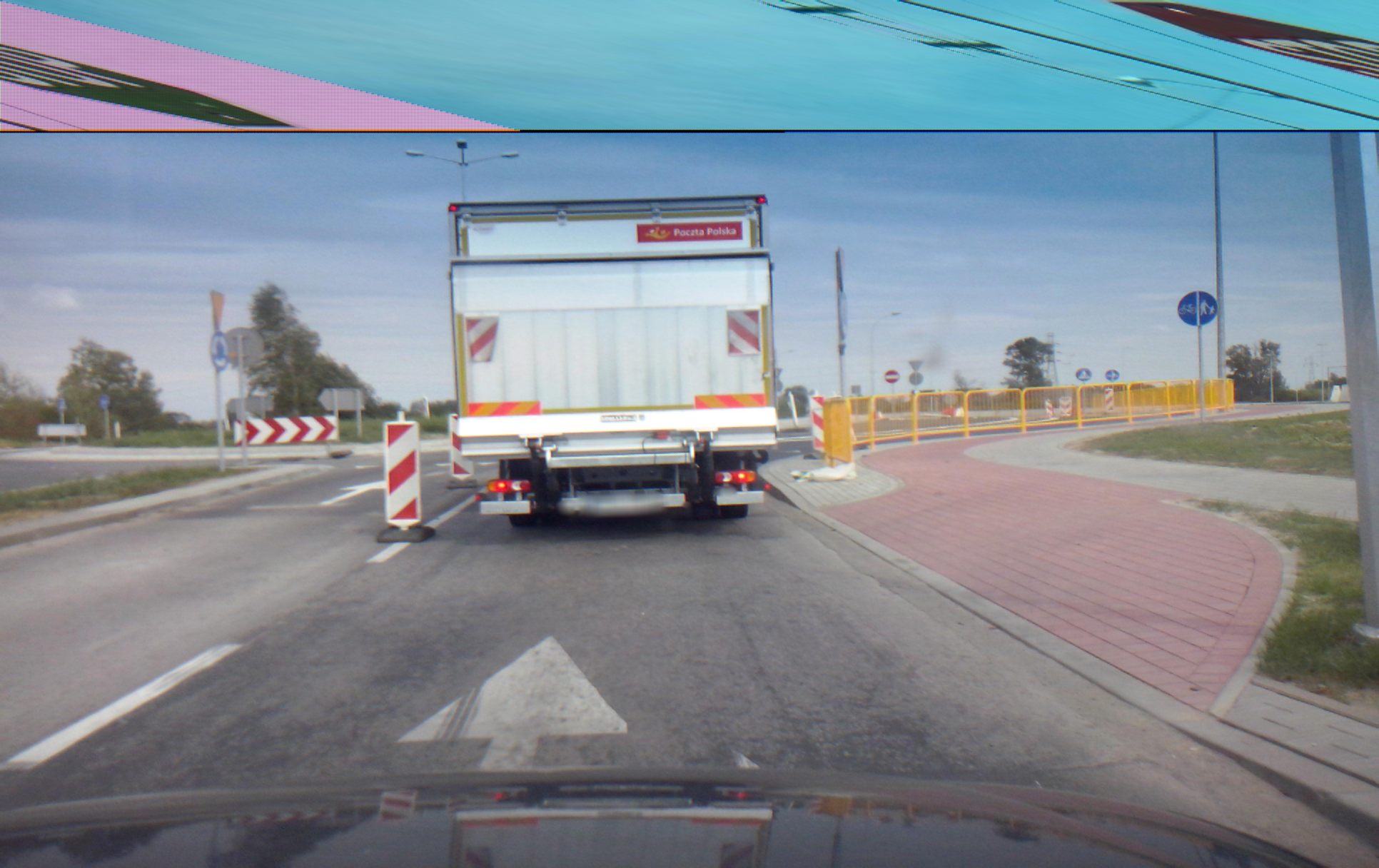}
    \caption{Width difference 4}
\end{subfigure}\vfill
\begin{subfigure}[t]{0.3\textwidth}
    \centering
    \includegraphics[width=\textwidth]{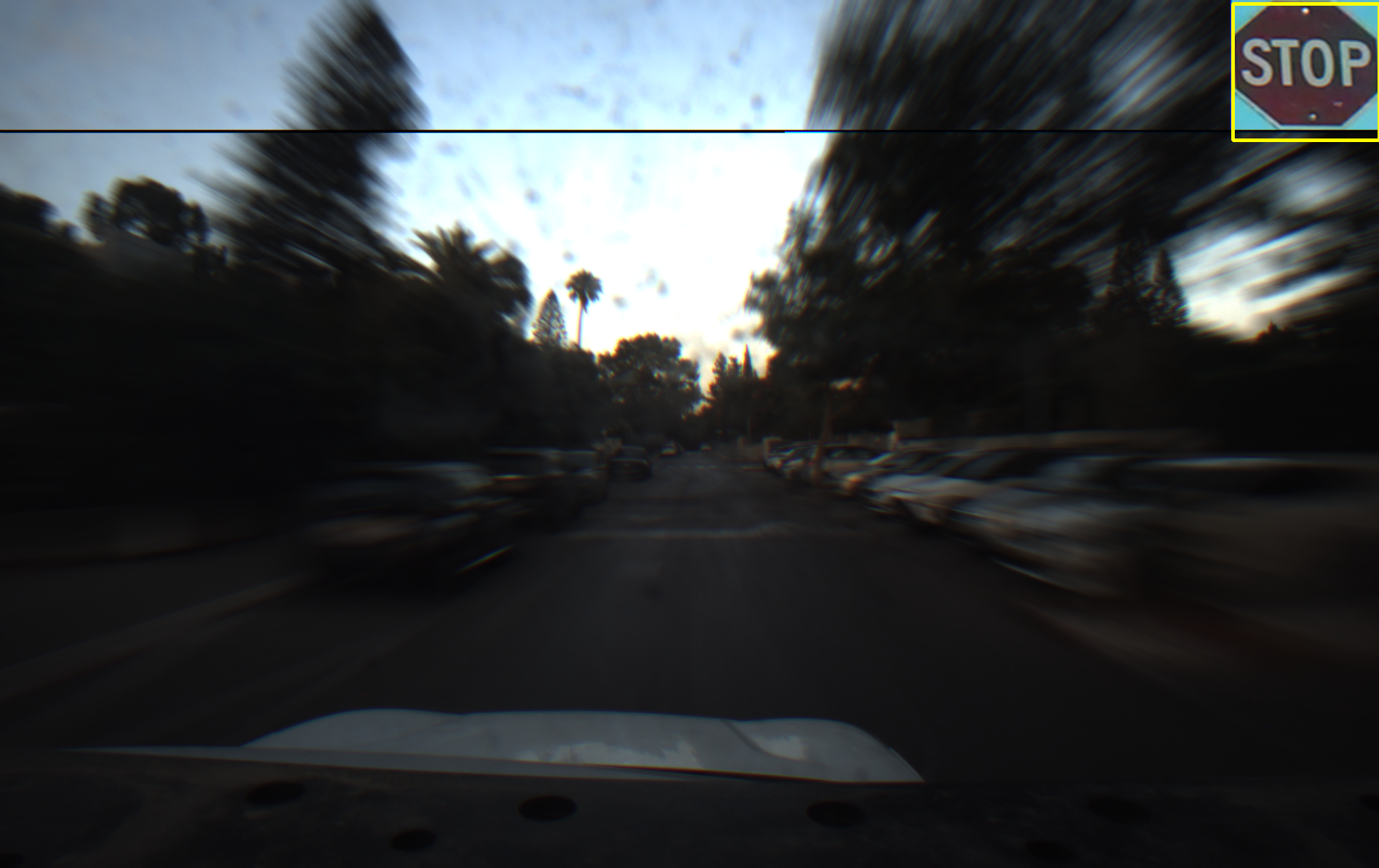}
    \caption{Width difference 0}
\end{subfigure}\hfill
\begin{subfigure}[t]{0.3\textwidth}
    \centering
    \includegraphics[width=\textwidth]{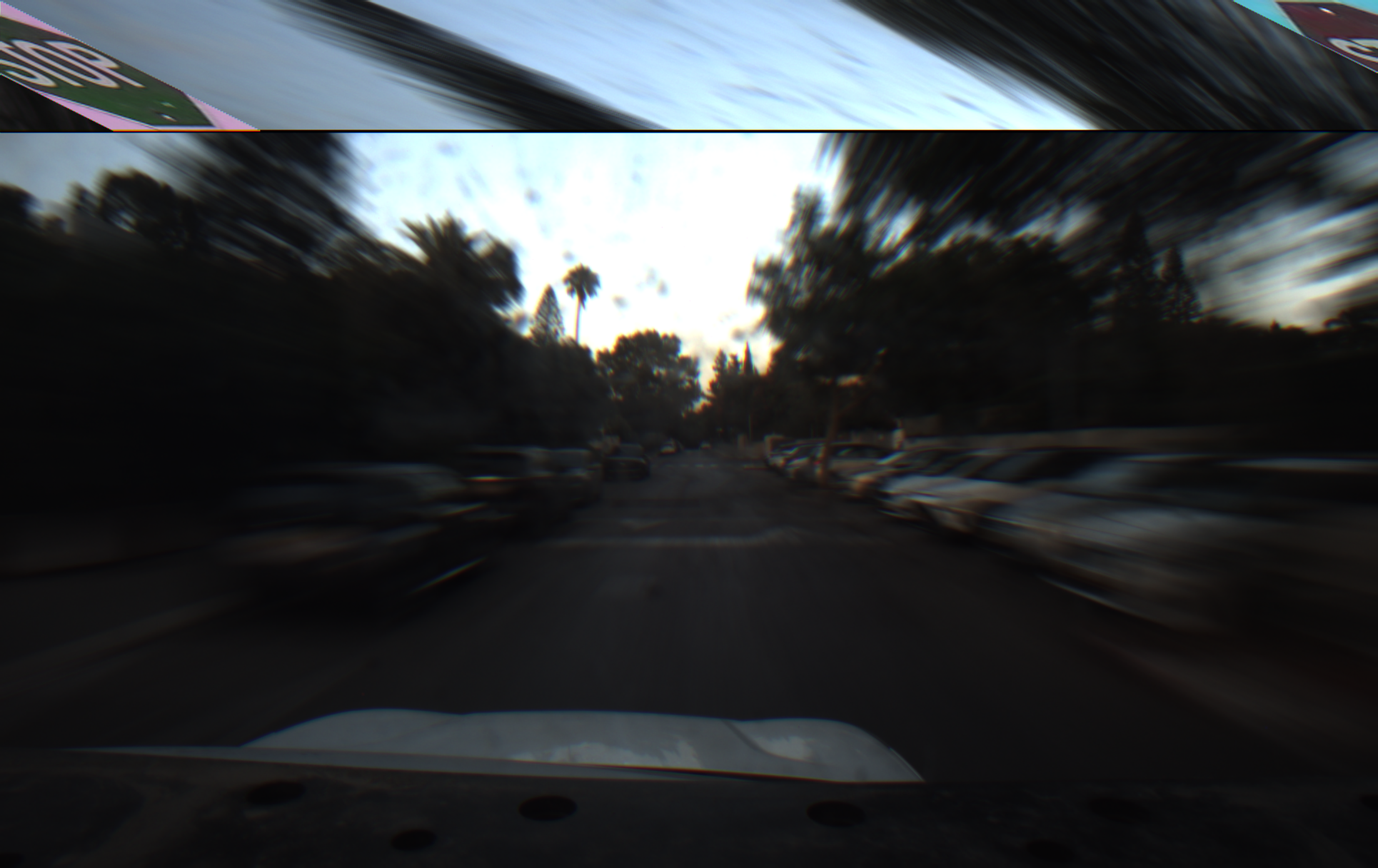}
    \caption{Width difference 2}
    \end{subfigure}\hfill
 \begin{subfigure}[t]{0.3\textwidth}
    \centering
    \includegraphics[width=\textwidth]{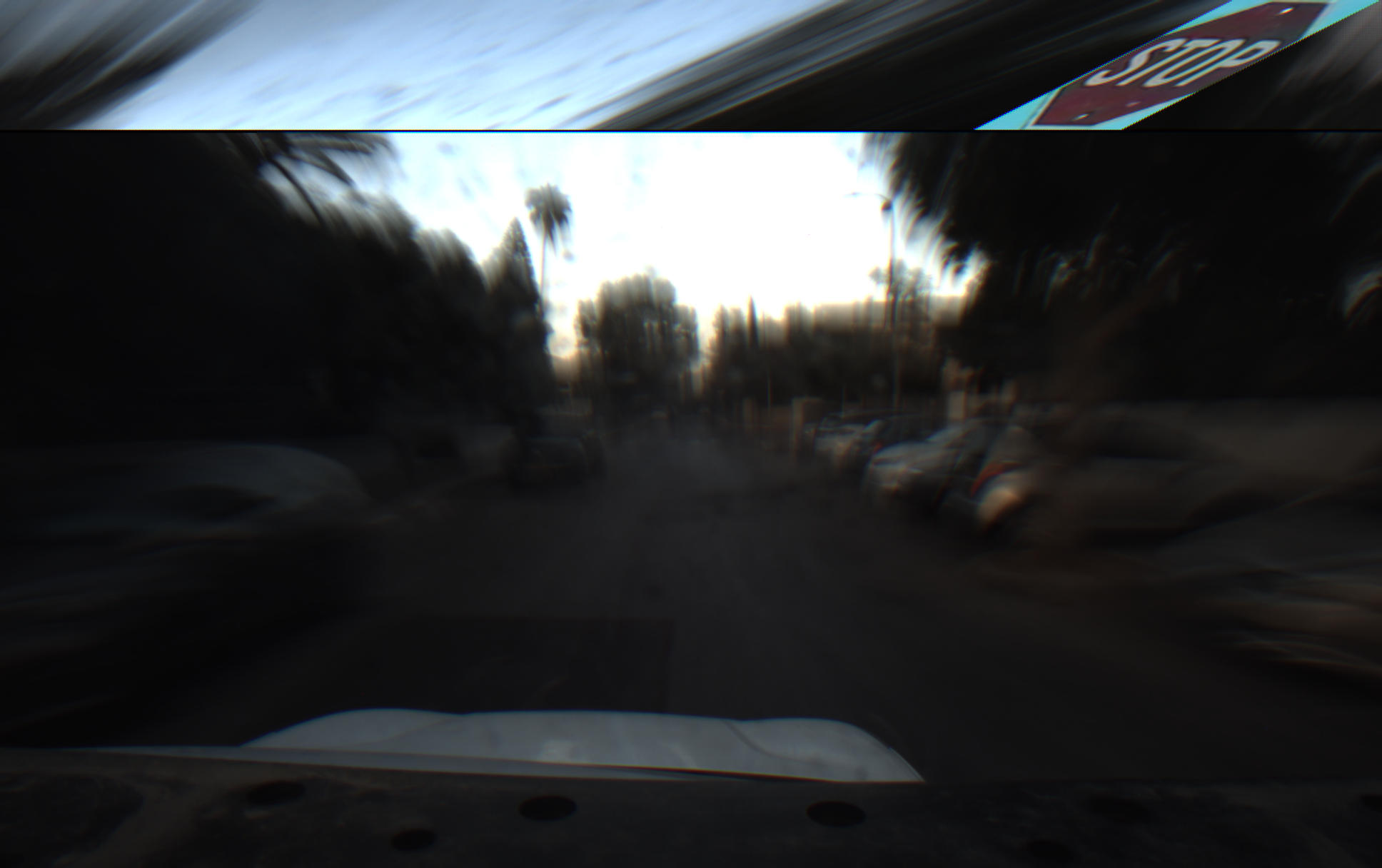}
    \caption{Width difference -2}
\end{subfigure}

\caption{Examples for stripe (a-c) and patch (d-f) injection attacks against the width-varying defense mechanism, with different difference between the injected width and true width. Note that only for the matching width injection in (a, d) the sign was recognized.
}
\label{fig:active_defense_stripe_injection_example}
\end{figure*}

\subsubsection{Example: A Basic attack on the Experimental Vehicle}

To demonstrate the behavior of our detectors, we describe their results against a basic, yet not totally naive, attacker: i.e., using the attack tool from Section~\ref{sec:attackTool} the attacker sniffs the real GVSP traffic, and uses the observed metadata to tweak the attack in real time. In the described scenario the attacker stops the camera, and repeatedly injects a fixed full-frame image of a stop sign.
The metadata of each fake frame includes a sequentially increasing frame ID, and a timestamp starting at 0 and incremented by 100 milliseconds, which matches the sample rate of the camera. The rate at which the packets were actually transmitted was higher.

The detection result of all the detectors in this full-frame attack (recall Section~\ref{sec:attackTool}) is depicted in Figure~\ref{fig:detections}. 

The top line in Figure~\ref{fig:detections} is the output of the traffic sign detector analyzing the video: we see that a (fake) stop sign is detected between IDs 50--100 and again between IDs 200--250. 

The ``Combined Detector'' line (red) is the logical ``OR'' of all the detectors, meaning that it is true if at least one of the detectors is true. Below it we see the indications of the 4~protocol-based detectors described in Section~\ref{sec:networkBased}: Constant Metadata, Frame ID, Timestamp, Timestamp Rate, and the 3 video-based detectors from Section~\ref{sec:videoBased}: MSE, Histogram, and Optical Flow.

The constant metadata detector (green) did not show any anomalies because we designed the fake frames to meet the expected parameters of the camera. 

The Frame ID detector (purple) exhibits a false alarm, due to a packet loss in the stream. In addition, it misses the start of the attack, because the fake frame is constructed with knowledge of the previous frame ID. It triggers an alert at the end of the attack because when the attacker re-starts the camera, the frame ID does not continue from the last ID of the last fake frame. 

The Timestamp-based detector (orange) triggers an alert at the beginning and end of the attack because the timestamp metadata was not updated correctly in the fake frames. The Timestamp Rate detector (cyan) detects the difference between the fake timestamp values in the metadata and the actual time difference between frames. 

As for the video-based detectors, the MSE detector (reddish-pink) triggers an alert for the entire attack as well, because all the fraud frames are identical. Both the Histogram-based detector (lime) and the Optical Flow detector (pink) trigger an alert in the start and the end of the attack because there is a large change in the scenery between the real and fake images. 

Many of our detectors, and certainly the Combined Detector, are able to successfully detect the attacks of this basic attacker. However, as we shall see in the quantitative evaluation in Section~\ref{sec:EvalPassiveDefense}, a non-naive attacker can evade them quite easily.

\subsection{Active Defense} \label{sec:ActiveDefense}
We now introduce a novel class of active defenses that embed randomly-generated data within the video stream in a manner challenging for attackers to manipulate, coupled with a verification process for correct decoding. This approach requires an external unit to control camera metadata or parameters through a dedicated channel and ensure decoding with a low false alarm rate. While our previous analysis in Section~\ref{sec:networkBased} treated metadata parameters like frame width, height and color encoding as constants, flagging deviations as anomalies, we now propose actively modifying these parameters with each new frame. The resulting changes can be easily decoded upon frame reception. Other possible parameters that an active defense can utilize include adjustable camera parameters like gain and exposure time. These adjustments cause noticeable changes in the video stream, such as variations in brightness, though verifying these changes is more complex compared to validation of metadata parameters.

A specific implementation of this active defense concept, which we call the ``width-varying defense'' involves dynamic modification of the \textit{width} of each frame to a randomly chosen width, using a GVCP command, thereby encoding a symbol of a few bits of data through the frame width. When a frame is received from the camera via GVSP, its received width is detected, and is verified against the randomized earlier-requested width.
As we shall see below, the width-varying defense has some extra advantages, and offers enhanced protection against certain types of injections.

\begin{figure}[t]
    \centering
\begin{subfigure}[t]{0.5\textwidth}
        \centering
        \resizebox{\textwidth}{!}{%
        \begin{circuitikz}
        \tikzstyle{every node}=[font=\normalsize]
        \draw [ line width=1pt ] (4,10.25) rectangle (5,1);
        \node [font=\LARGE] at (4.5,10.75) {Camera};
        \node [font=\LARGE] at (0.25,10.75) {Attacker};
        \draw [, line width=1pt ] (-0.5,10.25) rectangle (0.5,1);
        \draw [, line width=1pt ] (8.5,10.25) rectangle (9.5,1);
        \node [font=\LARGE] at (9,10.75) {Active Defense Unit};
        \node [font=\normalsize] at (6.75,9.5) {GVSP: Frame W};
        \draw [line width=1pt, ->, >=Stealth] (5,9.25) -- (8.5,9.25);
        \draw [line width=1pt, ->, >=Stealth] (0.5,5.25) -- (4,5.25);
        \node [font=\normalsize] at (1.75,5.5) {GVCP Stop};
        \draw [line width=1pt, ->, >=Stealth] (0.5,4.75) -- (8.5,4.75);
        \draw [line width=1pt, ->, >=Stealth] (5,6.25) -- (8.5,6.25);
        \draw [line width=1pt, ->, >=Stealth] (0.5,2) -- (4,2);
        \node [font=\normalsize] at (1.75,2.25) {GVCP Start};
        \draw [line width=1pt, ->, >=Stealth] (5,7.75) -- (8.5,7.75);
        \draw [, line width=1pt ] (13,10.25) rectangle (14,1);
        \node [font=\LARGE] at (13.5,10.75) {ADAS};
        \draw [line width=1pt, ->, >=Stealth] (9.5,9.25) -- (13,9.25);
        \draw [line width=1pt, ->, >=Stealth] (8.5,8.5) -- (5,8.5);
        \node [font=\normalsize] at (6.25,8.75) {GVCP: W1};
        \node [font=\normalsize] at (6.75,8) {GVSP: Frame W1};
        \node [font=\LARGE] at (7.25,8) {};
        \draw [line width=1pt, ->, >=Stealth] (8.5,7) -- (5,7);
        \node [font=\normalsize] at (6.25,7.25) {GVCP: W2};
        \node [font=\normalsize] at (6.75,6.5) {GVSP: Frame W2};
        \draw [line width=1pt, ->, >=Stealth] (8.5,5.5) -- (5,5.5);
        \node [font=\normalsize] at (6.25,5.75) {GVCP: W3};
        \draw [line width=1pt, ->, >=Stealth] (9.5,6.25) -- (13,6.25);
        \draw [line width=1pt, ->, >=Stealth] (9.5,7.75) -- (13,7.75);
        \node [font=\normalsize] at (11.25,8) {Valid Frame W1};
        \node [font=\normalsize] at (11.25,6.5) {Valid Frame W2};
        \node [font=\normalsize] at (2.25,5) {GVSP: Frame $W_{max}$};
        \draw [line width=1pt, ->, >=Stealth] (8.5,4) -- (5,4);
        \node [font=\normalsize] at (6.25,4.25) {GVCP: W4};
        \draw [line width=1pt, ->, >=Stealth] (0.5,3.25) -- (8.5,3.25);
        \node [font=\normalsize] at (2.25,3.5) {GVSP: Frame $W_{max}$};
        \draw [line width=1pt, ->, >=Stealth] (8.5,2.5) -- (5,2.5);
        \node [font=\normalsize] at (6.25,2.75) {GVCP: W5};
        \draw [line width=1pt, ->, >=Stealth] (5,1.75) -- (8.5,1.75);
        \node [font=\normalsize] at (6.75,2) {GVSP: Frame W5};
        \draw [line width=1pt, ->, >=Stealth] (9.5,4.75) -- (13,4.75);
        \node [font=\normalsize] at (11,5) {Invalid Frame};
        \node [font=\normalsize] at (11,3.5) {Invalid Frame};
        \draw [line width=1pt, ->, >=Stealth] (9.5,3.25) -- (13,3.25);
        \draw [line width=1pt, ->, >=Stealth] (9.5,1.75) -- (13,1.75);
        \node [font=\normalsize] at (11.25,2) {Valid Frame W5};
        \end{circuitikz}
        }%
    \end{subfigure}
\caption{Detection of full frame injection by the width-varying defense} \label{fig:active_defense_scheme}
\end{figure}
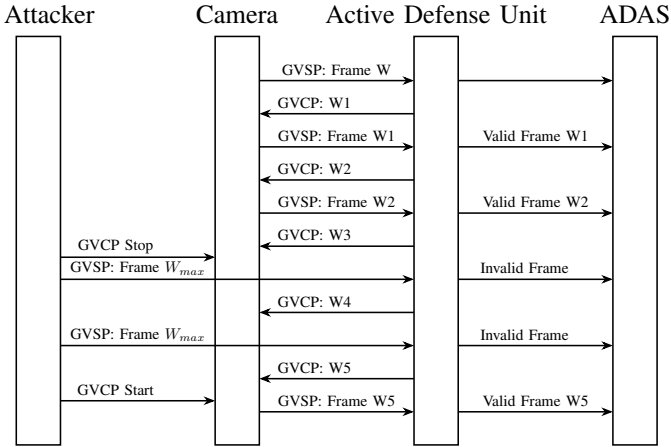

The width-varying defense functions as an anomaly detector against full-frame injection attacks: As we mentioned in the Adversary Model (Section~\ref{sec:AdversaryModel}) the real-time attacker can inject a predefined frame embedding a stop sign, but is unable to rebuild all the frame's packets on the fly to match a new width in real time. The intrusion detection is illustrated in Figure~\ref{fig:active_defense_scheme}.

Conversely, stripe injection and patch injection attacks would not trigger an anomaly, as they do not alter the frame width in the leader packet. However, injecting a stripe or patch with an incorrect width causes the stripe/patch image to undergo two transformations: diagonal spreading,  and color conversion as a result of the shift in the Bayer encoding. After these transformations, the traffic signs detector in the ADAS logic is likely to fail to recognize the injected sign: in other words, against such an attack the width-varying defense becomes an ``intrusion protection system'' rather than just an ``intrusion detection system''.

Also, in our experience, the small perturbations in the frame width are practically unnoticeable by humans even if the video stream is displayed on screen.

Figure~\ref{fig:active_defense_stripe_injection_example} illustrates the effectiveness of the method, showing that the traffic sign detector detects the stop sign only when injected with the correct width. The diagonal spreading direction is determined by the sign (negative or positive) of the width difference, as seen in (e) and (f). The color transformation is seen in (c) and (e) where the distorted stop sign changes its color to green.

The number of bits each symbol consists of is a parameter that involves a trade-off. On one hand, increasing the number of bits enhances the system's detection power because an attacker is less likely to inject an image with the correct width, making it easier to identify tampering. On the other hand, more bits-per-symbol results in more columns being omitted compared to a full-width frame, potentially reducing image quality or completeness. However, as we shall in the theoretical analysis (Section~\ref{sec:theoreticalAnalysis}), excellent detection probabilities can be achieved with as few as 1, 2, or 3 bits per symbol.

In Bayer encoding, the image's width must be a multiple of 2. Therefore, a single bit permits 2 symbols: the maximal width and a width 2 pixels less. Similarly, 2 bits allow for 4 symbols, ranging from the maximal width to a width 8 pixels less, and so:
\begin{equation}
Number Of Symbols = 2^b
\end{equation}
where \(b\) is the number of bits. The width \(W_k\) for a given symbol can be calculated as:
\begin{equation}
W_k = W_{max} -2k
\end{equation}
where $k$ is an integer between 0 and $2^b-1$. For example, with $b=2$ bits per symbol and $W_{max}=1936$ we have 4 symbols (00, 01, 10, 11), and 4 corresponding widths $W_k \in (1936, 1934, 1932, 1930)$.

To generate the required randomness we use the RC4 stream cipher after discarding the first 1000 pseudo-random output bytes.
The retrieved cipher bytes are flattened, meaning their concatenated bit representation is utilized. For each new frame, we extract from this representation the necessary number of bits based on the total number of possible widths. Subsequently, these bits are decoded into their integer representation. For instance, in a setup with $b=1$ bits and 2 symbols, a single RC4 byte suffices for 8 frames.

Since the defense mechanism is decoupled from the camera, the timing of sending the GVCP commands to change the camera's width is not perfectly synchronized with the frame rate, potentially resulting in delays in the frame width adjustment. Preliminary observations of our implementation indicated that the next detected width either matches the expected width, or matches the width requested for the previous frame. In other words we observed either no delay or a one-frame delay. Therefore, in order to make the decoder more robust, the width of the received frame is compared with the last two requested widths: if neither value matches, the frame is flagged as ``Invalid''.

\section{Evaluation}
\subsection{Passive Defense} \label{sec:EvalPassiveDefense}
We evaluated the video-based detectors (Histogram and Optical Flow) using two datasets: our own experimental dataset and the BDD100k dataset. In our experimental dataset, we analyzed 10,000 pairs of consecutive frames. The frames captured during these experiments have dimensions of 1936$\times$1216 pixels. The driving scenarios in our experimental dataset represent the urban environment around our university, and the daylight lighting and weather conditions in springtime.

Additionally, for each of 1000 videos selected from the BDD100k test dataset, we sampled 100 pairs of consecutive frames, resulting in a total of 100,000 unique instances. Frames in the BDD100k dataset were originally sized at 1280$\times$760 pixels and were resized to 1936$\times$1216 pixels using a bilinear interpolation to match the same dimensions as in our dataset. The BDD dataset is much more diverse than our experimental dataset, and includes videos from various scenes captured both during daytime and nighttime.

Each sequence of consecutive frames is used twice in our analysis. First, it is employed to establish a baseline score for a normal stream. Subsequently, to simulate an anomaly, one of the described attacks (full frame injection, stripe injection, or patch injection) is employed on the second frame of the pair.

We used two different fake images to test the injection, one with a stop sign and the other with a red light. For the stripe injection and the patch injection, we initially determined the minimum number of rows necessary for the MobileNet \cite{Howard2017} detector to reliably detect the injected object. For the stop sign image, it is 135 rows, which is approximately 10\% of the frame, while for the red light image only 65 rows are needed. 

\begin{figure*}[t]
\centering
    \begin{subfigure}[t]{0.3\textwidth}
        \centering
        \includegraphics[width=\textwidth]{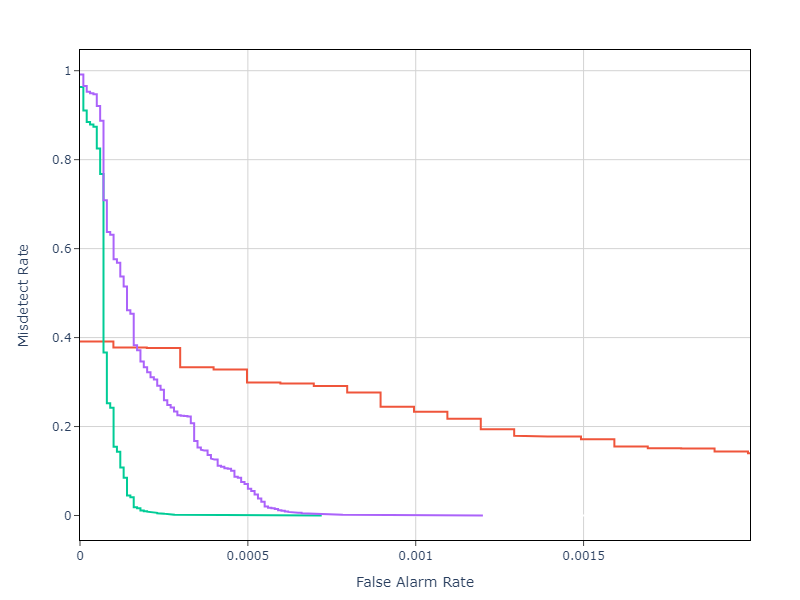}
        \caption{Full Frame Injection: Histogram}
    \end{subfigure}\hfill
    \begin{subfigure}[t]{0.3\textwidth}
        \centering
        \includegraphics[width=\textwidth]{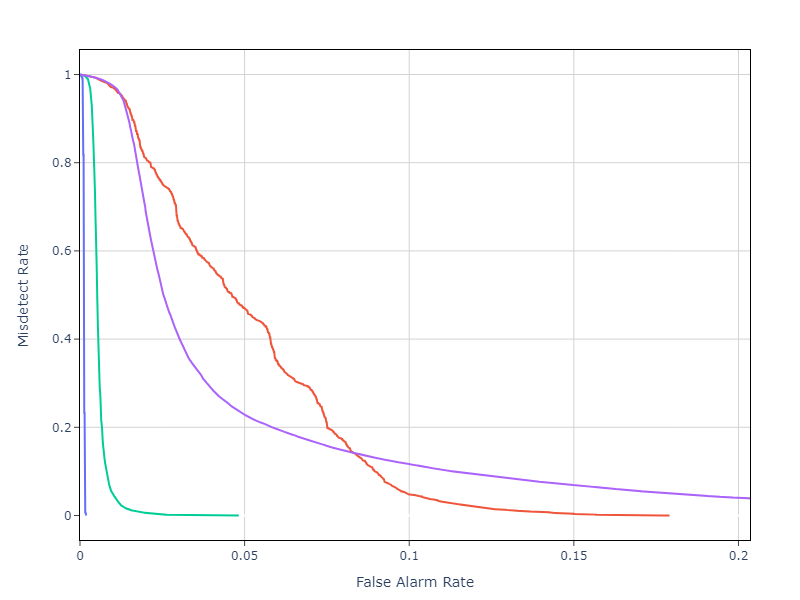}
        \caption{Stripe Injection: Histogram}
    \end{subfigure}\hfill
    \begin{subfigure}[t]{0.3\textwidth}
        \centering
        \includegraphics[width=\textwidth]{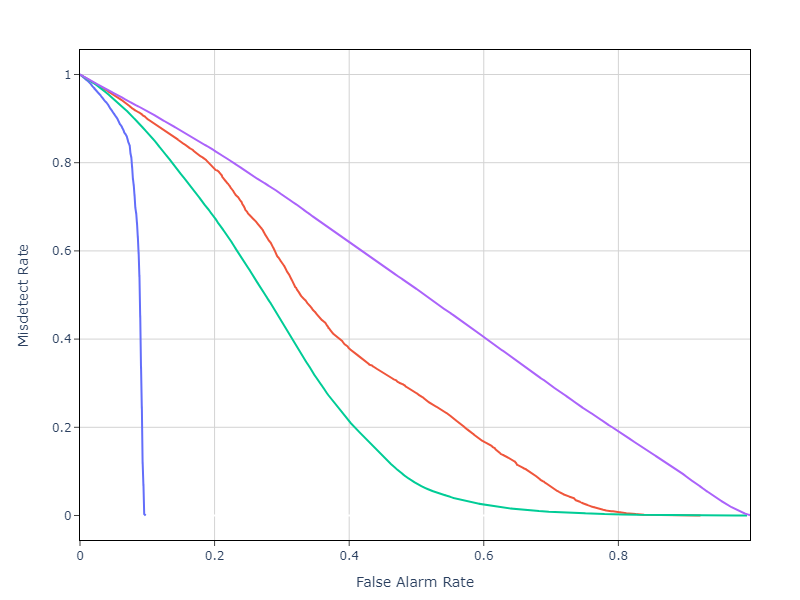}
        \caption{Patch Injection: Histogram}
    \end{subfigure}\vfill
    \begin{subfigure}[t]{0.3\textwidth}
        \centering
        \includegraphics[width=\textwidth]{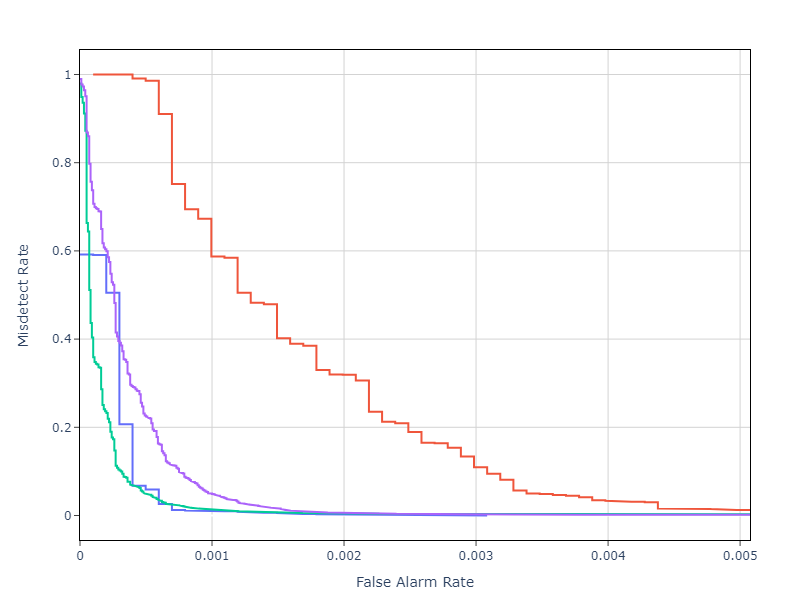}
        \caption{Full Frame Injection: Optical Flow}
    \end{subfigure}\hfill
    \begin{subfigure}[t]{0.3\textwidth}
        \centering
        \includegraphics[width=\textwidth]{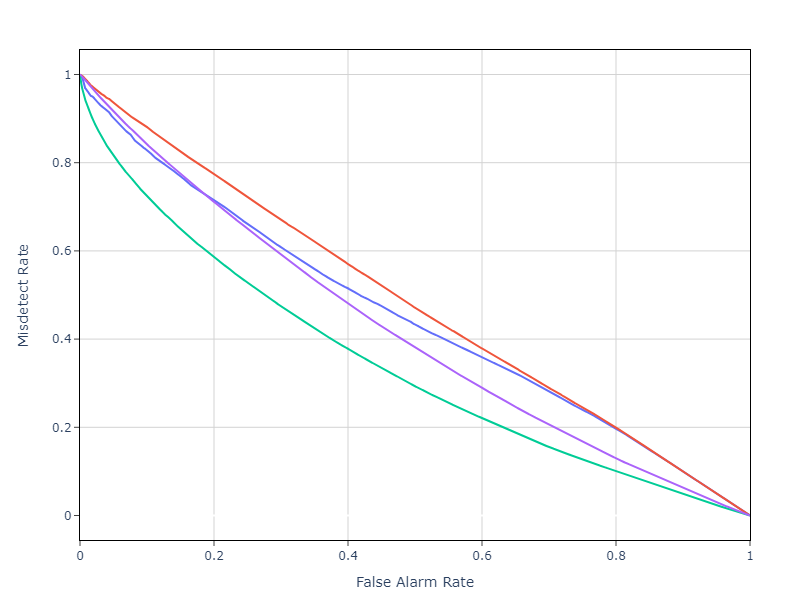}
        \caption{Stripe Injection: Optical Flow}
    \end{subfigure}\hfill
    \begin{subfigure}[t]{0.3\textwidth}
        \centering
        \includegraphics[width=\textwidth]{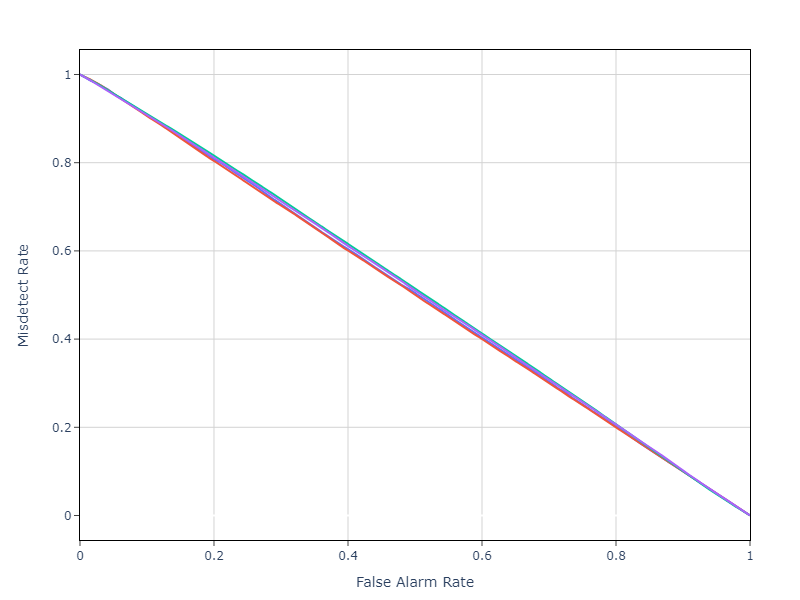}
        \caption{Patch Injection: Optical Flow}
    \end{subfigure}
\caption{DET curves for each detector are plotted under different manipulations. Blue and red lines represent evaluations of stop sign and red light objects, respectively, using the experimental dataset, while green and purple lines represent evaluations using the BDD100k dataset. Note that the X-axis range in (a), (b) and (d) is much smaller than in the other sub-figures.}
\label{fig:DET curves}
\end{figure*}

To evaluate the two methods, we used detection error tradeoff (DET) curve presented in Figure~\ref{fig:DET curves}for the two datasets. The misdetection rate (false negative rate) is plotted on the y-axis and the false positive rate (FPR) is on the x-axis. 

Both detectors exhibit robust performance in detecting the full frame injection attack, achieving an operating point marked by minimal false alarm and misdetection error rates. On the experimental dataset, the histogram-based detector's scores for the stop sign injection were in fact perfect, with zero misdetections or false alarms for thresholds ranging from 0.35 to 0.45. Consequently, its DET curve is not plotted. In the case of red light injection, which is a more stealthy form of injection, the misdetection rate decreases to values of less than 1\% only when the false alarm rate exceeds 5\%.

In the case of the stripe injection attack, see Figure~\ref{fig:DET curves}(e), the optical flow-based detector shows very poor performance on both datasets, as it is solely influenced by the small number of points-of-interest in the upper $\approx$10\% of the frame rows. Conversely, the histogram-based detector Figure~\ref{fig:DET curves}(b) has an acceptable operating point in this scenario. 

As for patch attack, see Figures~\ref{fig:DET curves}(c, f), both the Histogram and Optical Flow detectors are effectively useless when evaluated using the BDD dataset: they perform at the level of a random guess. This is since the stop sign and the red light patches are small enough that the histograms and optical-flow results on an unaltered frame and a manipulated frame are very similar. 
On the experimental dataset, the histogram-based detector is still capable of distinguishing between legitimate and frames injected with a stop sign, but with lower accuracy compared to the stripe injection scenario.

We conclude that while the passive detectors do provide some anomaly detection capabilities against naive attackers, slightly more sophisticated attackers that inject stripes or patches into previous frames can evade detection.

\begin{figure*}[t]
\centering
\begin{subfigure}[t]{0.45\textwidth}
\centering
\includegraphics[width=\textwidth]{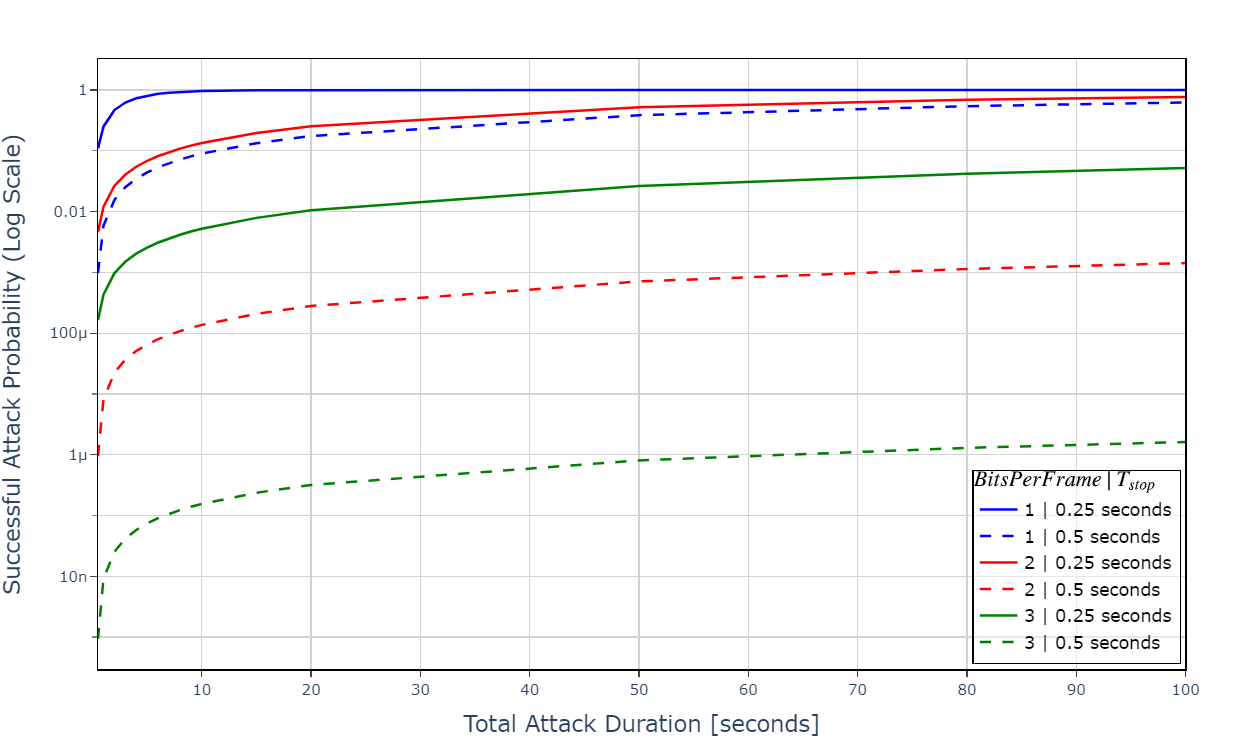}
\caption{Probability of a Successful Attack (log scale). The solid lines are for $T_{stop}= 0.25\,seconds$ and the dashed lines are for $T_{stop}= 0.5\,seconds$. }
\label{fig:successful_attack_probability}
\end{subfigure}\hspace{0.5cm}
\begin{subfigure}[t]{0.45\textwidth}
\centering
\includegraphics[width=\textwidth]{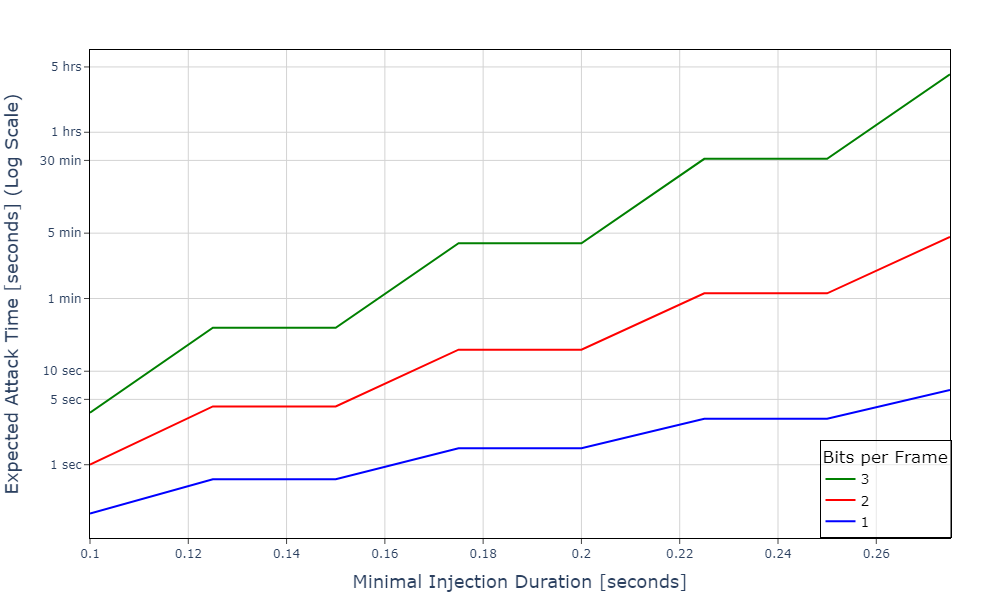}
\caption{Expected Waiting Time Until Successful Attack (log scale)
}

\label{fig:expected_waiting_time}
\end{subfigure}\hspace{0.5cm}
\caption{Theoretical Probability Evaluation of Injection Attacks}
\label{fig:probability_evaluation}
\end{figure*}

\subsection{The Width-Varying Defense}

\subsubsection{Theoretical Probability Evaluation} \label{sec:theoreticalAnalysis}
This section evaluates the effectiveness of the width-varying defense mechanism as an anomaly detector against full-frame injection attack, and as a protection mechanism against stripe and patch injection attacks, using a probabilistic approach. We analyze the likelihood of successful attacks and the expected waiting time until a successful attack occurs under various conditions.

Our evaluation is based on the following assumptions:
\begin{itemize}
\item The symbols of the pseudo-random sequence are uniformly distributed, resulting in a probability of a frame having a specific width given by:
\begin{equation}
P(W = w) = 1/2^b
\label{eq: width_probability}
\end{equation}
where $b$ is the number of bits transmitted per frame and $W$ is the width of the frame.
\item The traffic sign detector in the ADAS logic is perfect, always detecting traffic signs when they are injected with the correct width, and never detecting them otherwise.
\item In full-frame injection the attacker always uses a fixed width.
\item Different injection attempts are independent.
\item To stop the vehicle, the ADAS system must consistently detect the traffic sign over a sufficiently long time duration $T_{stop}$ \cite{nassi2020phantom}, which we call the ``minimal attack duration''. This duration translates to a requirement that the injected traffic sign is detected in a number of consecutive frames 
$N_{stop}=T_{stop} \times \mbox{\textit{fps}}$, where $\mbox{\textit{fps}}$ is the camera's frame-per-second rate. In our evaluation, we used a frame rate of 20Hz, and according to \cite{gillespie2021fundamentals} $T_{stop}$ needs to be in the range of [2.58, 5.25] seconds , implying $N_{stop} \in [52, 105]$.
\end{itemize}

For a single full-frame injection attack to evade our detection mechanism, the injected frame's width must match one of the $d_{max}+1$ most recently requested width values. The parameter $d_{max}$ reflects the maximum allowed delay in our detector (recall that in our implementation we used $d_{max}=1$). Therefore the probability of successfully detecting a single injected full-frame is given by:

\begin{equation}
P_{detection} = (1-1/2^b)^{d_{max}+1}.
\label{eq:detection_probability}
\end{equation}

The probability of a single successful stripe or patch injection equals the probability of injecting with the correct width, as shown in Equation~(\ref{eq: width_probability}), so the protection probability is:
\begin{equation}
P_{protection} = 1 - 1/2^b.
\label{eq:protection_probability}
\end{equation}
Let $p$ denote the probability of a successful injection (i.e., $p$ is either $1-P_{detection}$ or $1-P_{protection}$, depending on the attack variant). Then according to \cite{uspensky1937introduction, hald2005history}
the probability of having a run of at least $r$ consecutive successful injections within $n$ attempts is given by: 
\begin{equation}
P_{run}(n, r) = 1 - \beta_{n,r} + p^r\beta_{n-r, r},
\end{equation}
\begin{equation}
\beta_{n, r} = \sum_{\ell=0}^{n/(r+1)} (-1)^\ell \binom{n}{n-\ell r}(qp^r)^\ell,
\end{equation}
where $q = 1-p$.
Thus the probability of successful attack within $T_{attack}$ seconds is 
$P_{run}(T_{attack} \times \mbox{\textit{fps}} , N_{stop})$. 

Figure~\ref{fig:probability_evaluation}(a) illustrates the probability $P_{run}$ as a function of $T_{attack}$ for various numbers of bits per frame and $T_{stop}$ values. 
The figure shows that, as expected, the probability of successful attack decreases drastically for longer $T_{stop}$ values. Furthermore, using 3 bits per frame significantly reduces the probability of a successful attack.
Note that the graphs in Figure~\ref{fig:probability_evaluation}(a) paint an overly optimistic picture for the attacker: the curves are for $T_{stop}$ values an order of magnitude lower than the realistic range of [2.58, 5.25] seconds.

An alternate view of the adversary's likelihood of successfully stopping the vehicle involves analyzing the expected time it will take them to succeed: in other words, on average, how many images need to be injected until $r$ consecutive frames successfully evade detection, and how much time is this process expected to take.

Let $N_r$ denote the random variable counting the number of independent frame injection attempts until $r$ consecutive attempts evade detection, and let $T_r$ denote the time until this event happens. Then according to ~\cite{blom1993problems} the expectations of $N_r$ and $T_r$ are given by
\begin{equation}
E[N_r] = \sum_{\ell=0}^{r}\frac{1}{p^\ell},
\end{equation}
\begin{equation}
E[T_r] = E[N_r]/\mbox{\textit{fps}}
\end{equation}
where as before $p$ is the probability of a single injection to evade detection.
$E[T_r]$ is plotted in Figure~\ref{fig:probability_evaluation}(b) as a function of the minimal attack duration $T_{stop}$. Assuming that the attacker needs to have the stop sign recognized by the ADAS for at least $0.25$ seconds (i.e., $r=5$ at 
$\mbox{\textit{fps}}=20$Hz, which is much lower than the actual expected range), the figure shows that against a width-varying defense with 1-bit symbols the attacker will need to wait about 3 seconds until the attack succeeds---however already for 2-bit symbols they will need to wait $\approx 1$ minute, and for 3-bit symbols the wait time grows to 30--60 minutes. For a realistic value of, e.g., $T_{stop}=$2.58 seconds and $r=52$, with 1-bit symbols we get $E[T_r]\approx$ 15 million years, far outside the range of Figure~\ref{fig:probability_evaluation}(b).
These results demonstrate the effectiveness of our proposed protection mechanism in significantly reducing the likelihood of successful injection attacks on vehicular cameras.

\subsubsection{Experimental Evaluation: Data Collection}
Beyond the theoretical analysis we conducted real-world experiments to assess the effectiveness of our approaches under various conditions, using the experimental vehicle and our attack tool (recall Sections~\ref{sec:ExperimentalCar} and~\ref{sec:attackTool}).

During May and June 2024, we collected data using the experimental vehicle. For the proof of concept we configured the camera with a lowered frame rate of \(5 Hz\). The vehicle was driven in various environments, including urban and highway settings, during both day and night conditions. Each recording segment lasted 30 seconds, during which we captured the communication between the computer and the camera. 
Overall we captured 1.46 hours of driving video consisting of 26232 individual frames (41.19 GB of raw data).

To evaluate the real-time detection capability, we executed full-frame injection attacks in half of the recordings. After a few seconds from the start of each affected segment, we initiated a full-frame injection attack lasting for several seconds. This approach allowed us to test our detection mechanism's responsiveness and accuracy.

To assess the protection against stripe and patch injection attacks, we performed offline injections of traffic signs into the recorded frames. The injected signs were sourced from the Mapillary traffic sign dataset for detection \cite{ertler2020mapillary}. This method enabled us to evaluate our protection mechanism's effectiveness against a diverse range of potential injections.

\begin{table*}[t]
\centering
\begin{tabular}{|c|cc|cc|c|cc|cc|c|}
\hline
\multirow{3}{*}{\begin{tabular}[c]{@{}c@{}}Num Bits\\ Per Frame\end{tabular}} & \multicolumn{5}{c|}{Original} & \multicolumn{5}{c|}{Full Frame Injection} \\ \cline{2-11} 
 & \multicolumn{2}{c|}{Day} & \multicolumn{2}{c|}{Night} & \multirow{2}{*}{Total} & \multicolumn{2}{c|}{Day} & \multicolumn{2}{c|}{Night} & \multirow{2}{*}{Total} \\ \cline{2-5} \cline{7-10}
 & Urban & Highway & Urban & Highway & & Urban & Highway & Urban & Highway & \\ \hline
1 & 4739 & 499 & 2626 & 659 & \textbf{8523} & 735 & 298 & 645 & 86 & \textbf{1764} \\
2 & 4185 & 213 & 2055 & 454 & \textbf{6907} & 459 & 186 & 376 & 46 & \textbf{1067} \\
3 & 4115 & 146 & 2083 & 587 & \textbf{6931} & 489 & 121 & 404 & 26 & \textbf{1040} \\ \hline
\end{tabular}%
\caption{Number of frames collected under different conditions and bit rates}
\label{tab:frame_counts}
\end{table*}

\begin{table*}[t]
\centering
\begin{tabular}{|c|cc|cc|c|}
\hline
\multirow{2}{*}{\begin{tabular}[c]{@{}c@{}}Num Bits\\ Per Frame\end{tabular}} & \multicolumn{2}{c|}{Day} & \multicolumn{2}{c|}{Night} & \multirow{2}{*}{Total} \\ \cline{2-5}
 & Urban & Highway & Urban & Highway & \\ \hline
1 & 56 (11.08GB) & 11 (1.94 GB) & 36 (4.26 GB) & 8 (321.9 MB) & 111 (17.6 GB) \\
2 & 46 (8.2 GB) & 8 (836.12 MB) & 24 (2.74 GB) & 4 (115.66 MB) & 83 (11.87 GB) \\
3 & 44 (8.98 GB) & 5 (367.04 MB) & 22 (2.21 GB) & 6 (171.08 MB) & 77 (11.72 GB) \\ \hline
\end{tabular}
\caption{Number of 30-second captures, and recording sizes, under different conditions and bit rates.}
\label{tab:files_count}
\end{table*}

For each recording, we varied the number of transmitted bits per symbol to analyze the impact of this parameter on both detection and protection performance. Table~\ref{tab:frame_counts} summarizes the number of collected frames under different conditions and bit rates, and Table~\ref{tab:files_count} summarizes the total number of experiments. The frames counted under ``Original'' were used to evaluate the protection mechanism, while those counted under ``Full Frame Injection'' were used to assess the detection mechanism.

\subsubsection{Detection Evaluation}
\begin{table*}[t]
\centering
\begin{tabular}{|c|ccc|ccc|ccc|ccc|}
\hline
\multirow{3}{*}{\begin{tabular}[c]{@{}c@{}}Num Bits\\ Per Frame\end{tabular}} & \multicolumn{6}{c|}{Day} & \multicolumn{6}{c|}{Night} \\ \cline{2-13} 
 & \multicolumn{3}{c|}{Urban} & \multicolumn{3}{c|}{Highway} & \multicolumn{3}{c|}{Urban} & \multicolumn{3}{c|}{Highway} \\ \cline{2-13} 
 & Valid & Invalid & \begin{tabular}[c]{@{}c@{}}Detection\\ Probability\end{tabular} & Valid & Invalid & \begin{tabular}[c]{@{}c@{}}Detection\\ Probability\end{tabular} & Valid & Invalid & \begin{tabular}[c]{@{}c@{}}Detection\\ Probability\end{tabular} & Valid & Invalid & \begin{tabular}[c]{@{}c@{}}Detection\\ Probability\end{tabular} \\ \hline
1 & 616 & 119 & 0.16 & 245 & 53 & 0.18 & 539 & 106 & 0.16 & 70 & 16 & 0.19 \\
2 & 261 & 198 & 0.43 & 104 & 82 & 0.44 & 213 & 163 & 0.43 & 28 & 18 & 0.39 \\
3 & 133 & 356 & 0.73 & 34 & 87 & 0.72 & 103 & 301 & 0.75 & 8 & 18 & 0.69 \\ \hline
\end{tabular}%
\caption{Detection probabilities across different conditions and bit rates}
\label{tab:detection_probabilities}
\end{table*}

\begin{figure}[t]
\centerline{\includegraphics[width=0.5\textwidth]{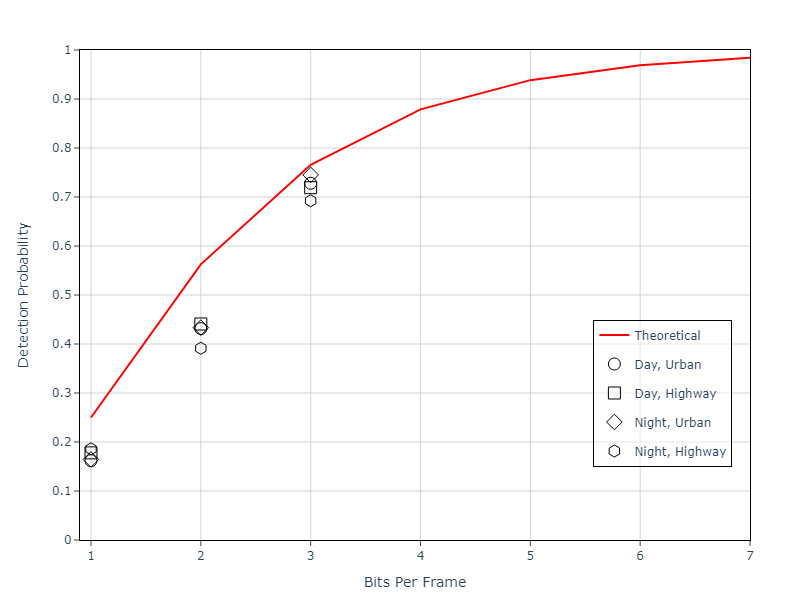}}
\caption{Detection probability of full-frame injection attack as a function of number of transmitted bits per frame. The red line is the theoretical probability $P_{detection}$ and the symbols present the measured values in different driving scenarios.}
\label{fig:experiments_evaluation_detection}
\end{figure}

During the full-frame attack injection, the attack tool injected frames with a constant width of 1936 pixels, the maximal width. Our detector, using a maximal delay of $d_{max}=1$, marks a frame as ``Invalid'' if the received width does not match one of the last two transmitted width values. Figure~\ref{fig:experiments_evaluation_detection} and Table~\ref{tab:detection_probabilities} present the measured detection probability of a single injected frame across various driving conditions and bit rates, alongside the theoretical probability curve derived from Equation~\ref{eq:detection_probability}.

The figure demonstrates a clear correlation between the number of bits per frame and the single-frame detection probability $P_{detection}$ from Equation~(\ref{eq:detection_probability}). As the bits per frame increase from 1 to 3, we observe a significant rise in detection probability from approximately 20\% to 70\%. The experimental results closely align with the theoretical curve, validating our model's accuracy.

The figure also shows that the various driving scenarios have minimal impact on the detection probabilities, as evidenced by the clustering of data points for each bit rate. This consistency across diverse conditions highlights the robustness of our detection mechanism.

The graph also suggests that increasing the bits per frame beyond 3 would yield diminishing returns in detection probability improvement, as the theoretical curve begins to flatten. This insight can guide optimal configuration of our system, balancing detection effectiveness with computational efficiency.

\begin{figure}[t]
\centerline{\includegraphics[width=0.5\textwidth]{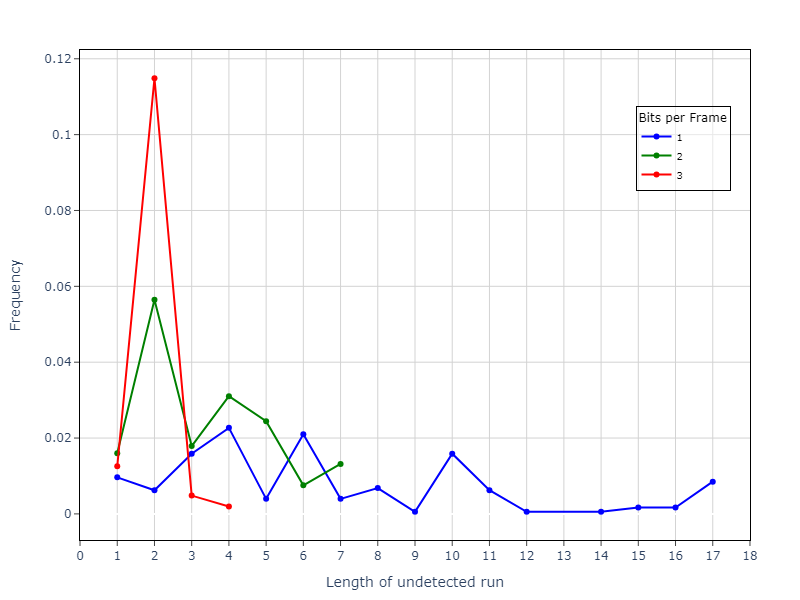}}
\caption{Distribution of runs lengths of consecutive undetected full frame injections, excluding runs of length 0 (of just invalid frames).}
\label{fig:undetected_run_length}
\end{figure}

Figure~\ref{fig:undetected_run_length} illustrates the distribution of runs lengths of consecutive undetected full-frame injections, across the entire dataset, for varying numbers of bits per frame. 
As the number of bits per frame increases, the likelihood of longer undetected sequences diminishes. Notably, sequences of length 2 occur most frequently, which can be attributed to the detector's window size. When extrapolating these results to a frame rate of $\mbox{\textit{fps}}=20Hz$ (although the original data was captured at a lower rate), the analysis using 3 bits per frame 
shows that the longest undetected injection lasted 0.2 seconds, well below the required $T_{stop}$ of 2.58 seconds, and such runs occurred with a 0.19\% probability. 
This demonstrates the effectiveness of the detection mechanism in limiting the duration of successful attacks, particularly at higher bit rates.

These findings not only confirm the effectiveness of our detection mechanism but also demonstrate its reliability across various real-world driving conditions, closely matching our theoretical predictions.

\begin{figure*}[t]
\centering
\begin{subfigure}[t]{0.45\textwidth}
\centering
\includegraphics[width=\textwidth]{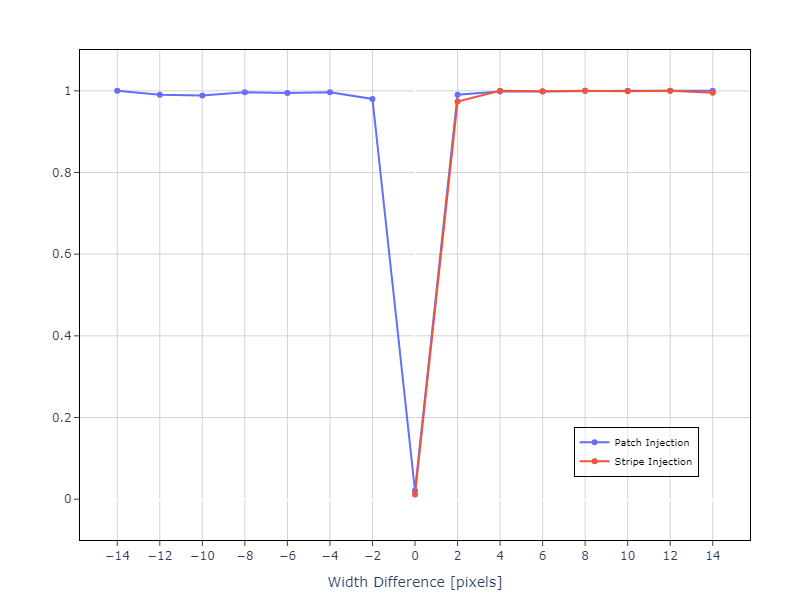}
\caption{Defense rate as a function of the difference between the injected width and the frame width.}
\label{fig:active_evaluation_width_diff}
\end{subfigure}\hspace{0.5cm}
\begin{subfigure}[t]{0.45\textwidth}
\centering
\includegraphics[width=\textwidth]{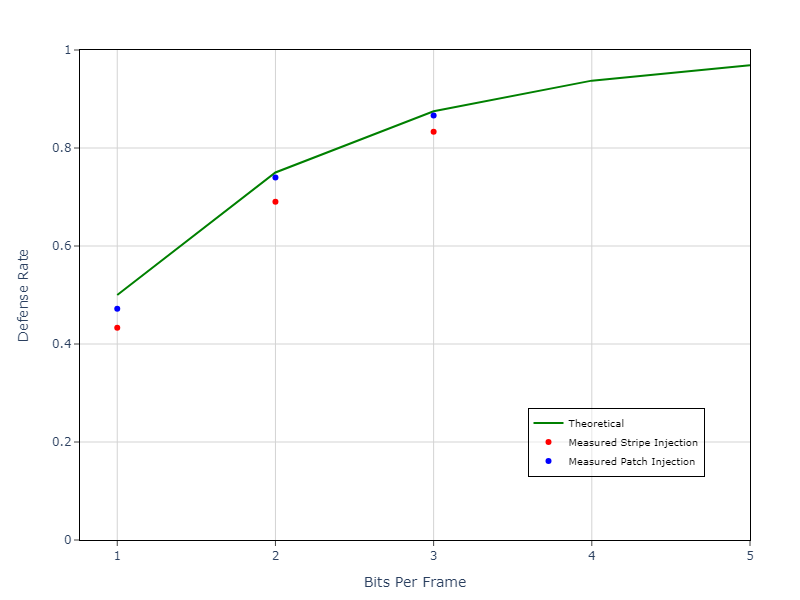}
\caption{Defense rate as a function of the number of bits transmitted with every frame.}
\label{fig:active_evaluation_total_defense}
\end{subfigure}\hspace{0.5cm}
\caption{Defense rate of the active mechanism}
\label{fig:defense_rate}
\end{figure*}

\begin{figure*}[t]
\centering
\begin{subfigure}[t]{0.45\textwidth}
\centering
\includegraphics[width=\textwidth]{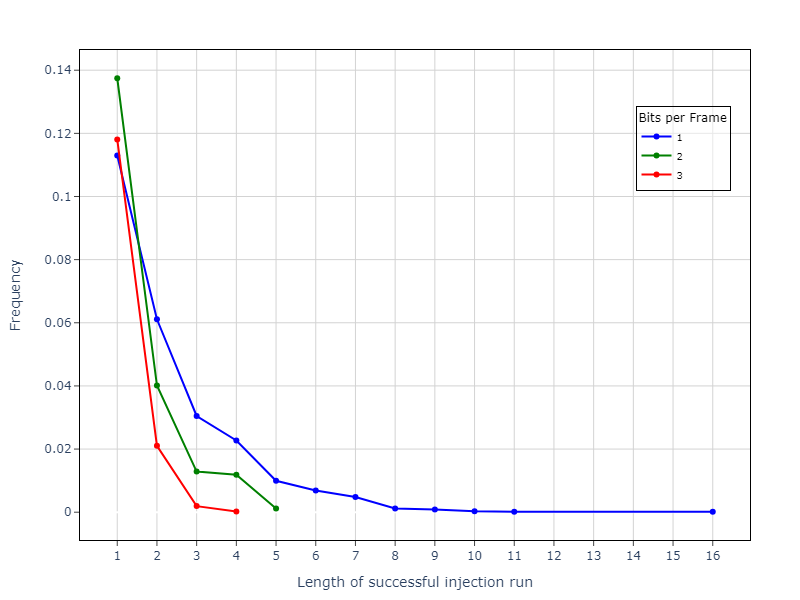}
\caption{Stripe Injection}
\label{fig:sucessful_run_length_stripe}
\end{subfigure}\hspace{0.5cm}
\begin{subfigure}[t]{0.45\textwidth}
\centering
\includegraphics[width=\textwidth]{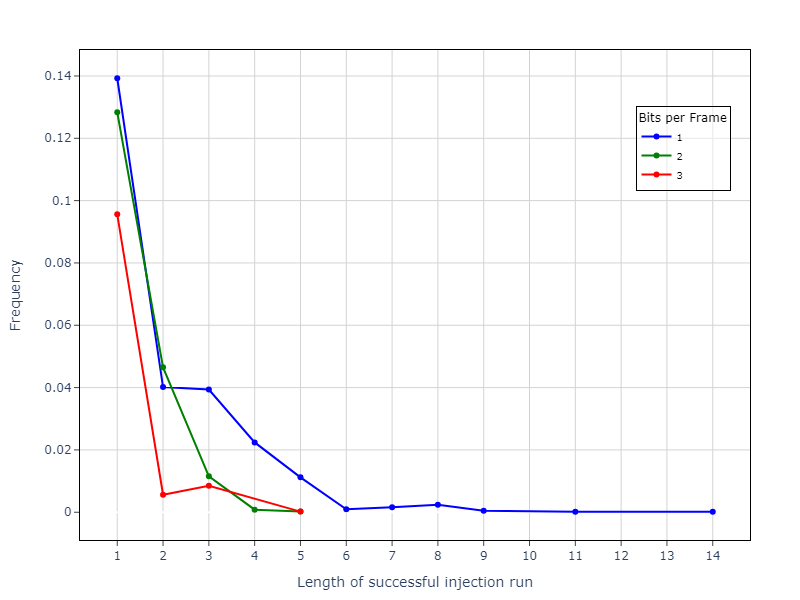}
\caption{Patch Injection}
\label{fig:sucessful_run_length_patch}
\end{subfigure}\hspace{0.5cm}
\caption{Distribution of runs lengths of successful consecutive stripe and patch injections in which MobileNet recognized the traffic sign}
\label{fig:sucessful_run_length}
\end{figure*}

\subsubsection{Protection Evaluation}
We also assessed the effectiveness of our width-varying defense as a protection mechanism against stripe and patch attacks. We used original frames from our corpus, and injected them with 204 stop sign images from the Mapillary traffic sign dataset \cite{ertler2020mapillary}. The traffic signs we selected are all successfully detected by MobileNet. Our goal was to quantify the defense's ability to degrade MobileNet's traffic sign detection after stripes or patches 
are injected into our captured frames, and distorted when there is a width mismatch.

We evaluated the stripe injection attack as follows:
\begin{itemize}
\item For every frame in our corpus, we randomly selected an image from the 204 Mapillary dataset images.
\item The selected image was resized to match the camera's maximum width of 1936 pixels.
\item We injected the stripe containing the stop sign into the upper part of each frame.
\end{itemize}

We evaluated the patch injection attack as follows:
\begin{itemize}
    \item We used the frames for which the preceding frame exists in the recording, plus those for which the difference in IDs was 2 (simulating single frame loss).
    \item For each such frame we randomly selected an image from the 204 images in the Mapillary dataset.
    \item We extracted a stripe from the top of the image of the prior frame, embedded a patch containing the stop sign into the stripe's top-right corner, and injected the stripe into the current frame.
\end{itemize}

For every injection, we ran MobileNet inference using sliding window of $484 \times 304$ and a confidence threshold of $0.3$, calibrated using the traffic sign dataset. We saved the highest confidence score of a bounding box intersecting with the injected area. We counted a defense event if the highest detection score was lower than the $0.3$ threshold.

Figure~\ref{fig:defense_rate}(a) shows the defense rate for both types of injections as a function of the difference between the injected width and the original width. Due to the Bayer encoding, only even differences are possible. For stripe injection (red line), the injected width was always the maximum, so only positive width differences are possible. In contrast, for patch injection, the width difference is between the widths of two consecutive frames, allowing for both positive and negative differences. The experiments utilized a maximum of 3 bits per frame, leading to 15 possible width differences between -14 and 14, with 0 indicating two consecutive frames with the same width. The defense rate against successful injections with a width difference of 0 is low as expected, however for injections where there is a width mismatch between the prior and current frame the defense rate is nearly optimal. Most failed defense events occurred at the minimal width difference of 2, where the traffic sign distortion is least destructive.

Figure~\ref{fig:defense_rate}(b) illustrates the total defense rate as a function of the number of bits transmitted per frame. The measured values correlate with the theoretical values given by Equation~\ref{eq:protection_probability}. For patch injection, the difference between theoretical and measured values is lower due to a higher misdetect rate of the object detector for successful injections.

Figure~\ref{fig:sucessful_run_length} illustrates the distribution of runs of consecutive successful injections,  where a stop sign is detected in each frame of the sequence. The graphs show that longer successful runs become less frequent as the number of bits per frame increases. Adjusted for $\mbox{\textit{fps}}=20Hz$, using 3 bits per frame, the stripe injection analysis over a total period of 259 seconds revealed a maximum successful duration of 0.15 seconds, occurring with a 0.19\% probability. Similarly, for patch injection over a 240 seconds period, the longest successful sequence also lasted 0.15 seconds but with a higher probability of 0.85\%.

This comprehensive evaluation demonstrates the effectiveness of our protection mechanism across various injection scenarios and bit rates. 

\section{Limitations and Future Work}
While the width-varying defense is quite effective against real-time attackers that can evade the passive protocol-based and video-based detectors, an even more sophisticated attacker, that is aware of the width-varying defense, can defeat it. Such an attacker can keep a repertoire of $2^b$ frames/stripes/patches, one per potential width. To defeat the defense an adaptive attacker needs to listen to the current frame's width request on the GVCP channel, and use the frame/stripe/patch suitable to the requested width. Nonetheless, we argue that the width-varying defense is a significant barrier against real-time attackers, and its simplicity makes it attractive.

We considered an extension of the width-varying defense, which simultaneously manipulates the image height and width. This effectively increases the number of random bits per frame, and improves the defense against a full-frame injection even further---but does not offer additional protection against stripe and patch injections. Since the 3-bit width-varying defense is already extremely effective against full-frame injection, we did not explore this further. 

We leave for future work further research into these techniques, and in particular exploring additional active defense mechanisms within the proposed class. Beyond the image dimensions we used, one can devise active defenses which involve manipulating other parameters such as image encoding, embedding of GPS information, or camera sensor parameters such as exposure time.

A parallel research direction is to evaluate active defense, and in particular the width-varying defense, on other camera types and communication protocols. While our current implementation focused on raw video streams, it would be valuable to adapt these techniques to compressed video formats, such as MPEG \cite{le1991mpeg} and H.264 \cite{wiegand2003overview}.

Finally, a more comprehensive defense would be to encapsulate the communication between the ADAS logic and the camera in a cryptographically-authenticated channel, such as a TLS connection or IPsec tunnel. However, we argue that using such technology is significantly more complex than the mechanisms we propose.

\section{Conclusions}
Protecting ADAS systems from attacks is a crucial aspect of vehicular defense. In this paper we demonstrated a new attack vector: manipulating the IP-based network communication between the camera and the ADAS logic, injecting fake images of stop signs or red lights into the video stream, and letting the ADAS stop the car safely. We created such an attack tool that successfully exploits the GigE Vision protocol. 

Then we analyzed two classes of passive anomaly detectors to identify such attacks: protocol-based detectors and video-based detectors. Our results show that such detectors are effective against naive adversaries, but sophisticated adversaries can evade detection.

Our best proposal is a novel class of active defense mechanisms that randomly 
adjust camera parameters during the video transmission, and verify that the received images obey the requested adjustments. Within this class we focused on the width-varying defense, which randomly modifies the width of every frame. Beyond its function as an anomaly detector, this defense is also a protective measure against certain attacks: by distorting injected image patches it prevents their recognition by the ADAS logic. Through detailed analysis and extensive evaluation we demonstrated its effectiveness against full-frame, stripe and patch injections. We believe that as ADAS and autonomous vehicles proliferate and are subject to more attacks, our active defense mechanism can play a significant defensive role.

\bibliography{citations.bib}{}
\bibliographystyle{IEEEtranS}

\end{document}